\documentclass[10pt,twoside]{IEEEtran}
\usepackage[utf8]{inputenc}
\usepackage[T1]{fontenc}
\usepackage{amssymb}
\usepackage{amsmath}
\usepackage{amsthm}
\usepackage{mathrsfs}
\usepackage{graphicx}
\usepackage{array}
\usepackage{tabularx}
\usepackage{xcolor}
\usepackage{cite}
\usepackage{psfrag}
\usepackage{subcaption}
\usepackage{booktabs}
\usepackage{multirow}
\usepackage{url}
\usepackage{siunitx}
\DeclareSIUnit{\belmilliwatt}{Bm}
\graphicspath{ {./fig/} }
\newtheorem{theorem}{Theorem}
\newtheorem{lemma}{Lemma}

\newtheorem{remark}{Remark}
\newtheorem{definition}{Definition}

\usepackage{color,soulutf8}
\definecolor{laranja}{rgb}{1.0, 0.6, 0.2}

\begin{document}

\title{Network-Coded Cooperative LoRa Network with D2D Communication}
\author{Luis Henrique de Oliveira Alves,
Jo\~{a}o~Luiz~Rebelatto,~\IEEEmembership{Senior~Member,~IEEE,}
Richard~Demo~Souza,~\IEEEmembership{Senior~Member,~IEEE}
and~Glauber~Brante,~\IEEEmembership{Senior~Member,~IEEE}
\thanks{L. H. O. Alves, J. L. Rebelatto and G. Brante, CPGEI/UTFPR, Curitiba-PR, Brazil, E-mails: luishoalves@gmail.com, \{jlrebelatto,gbrante\}@utfpr.edu.br}
\thanks{R. D. Souza, EEL/UFSC, Florian\'{o}polis-SC, Brazil, E-mail: richard.demo@ufsc.br.}
\thanks{This work has been partially supported by CNPq and CAPES (Brazil).}
}

\maketitle

\begin{abstract}
We evaluate the outage probability and the energy efficiency (EE) of a LoRa network when the end-devices (EDs) are capable of exchanging messages with a device-to-device (D2D) technology. Our proposal is to assess the performance of the network when operating under the  concept of network-coded cooperation (NCC), where the EDs are capable of transmitting linear combinations of more than one frame (including frames of another ED), performed over a non-binary finite field. We consider both connection and collision probabilities when modelling the outage probability of an ED, as well as a realistic power consumption model to evaluate the EE. Our results indicate that the proposed scheme, which we refer to as NCC-LoRa, provides considerable gains in terms of both outage probability and EE when compared to a regular LoRa network, even when considering the additional consumption of D2D communication.  
\end{abstract}

\begin{IEEEkeywords}
Cooperative communications, LoRa, LPWAN, network coding.
\end{IEEEkeywords}

\section{Introduction} \label{sec:introduction}

The advent of the Internet-of-Things (IoT) has generated interest from the industry and research community towards Low-Power Wide-Area Networks (LPWAN)~\cite{Centenaro2016,Raza2017}. A LPWAN is designed to support low cost applications that are tolerant to low transmission rates, but require low energy consumption. In this context, the proprietary Long Range (LoRa) technology has been developed~\cite{Ikpehai2019,Sundaram2020}. This technique is derived from Chirp Spread Spectrum (CSS) and is patented by Semtech~\cite{LoRaPATENT}.

Network implementations with LoRa are generally based on the LoRa Wide-Area Network (LoRaWAN) protocol, developed by the LoRa Alliance~\cite{LoRaWAN}. The multiple access in LoRaWAN is based on ALOHA, without any collision avoidance mechanisms~\cite{Sundaram2020}. With the prospect of IoT ubiquity and increase in network densities, system modelling~\cite{Georgiou2017,Mahmood2019} and methods aiming at improving performance have gained relevance~\cite{Hoeller2018,SantAna2020,SantAna2020SIC}.

As in~\cite{Georgiou2017,Mahmood2019,Hoeller2018,SantAna2020,SantAna2020SIC}, LoRaWAN typically operates in a star topology, where the End-Devices (EDs) communicate directly to the GateWay (GW), without inter-ED message exchanges. Although this is in line with the LPWAN low-complexity requirement, this approach may be less efficient in many scenarios, being Device-to-Device (D2D) communication an alternative to boost the network performance~\cite{Jiang2021,Hsu2018,Borkotoky2019,Mikhaylov2017,Kim2018}. 

Thus, one could expect the presence of D2D in future LoRaWAN implementations or even in the specification, not only to exchange information between a pair of EDs, but also enabling the EDs to cooperate towards transmitting their individual messages to the GW. The introduction of cooperative communication dates back to early 2000's~\cite{Laneman2004}. The technique is capable of emulating the effects of multiple antennas and achieving spatial diversity in a network composed of single-antenna EDs. The cooperative concept takes advantage of the broadcast nature of the wireless channel, by allowing EDs to receive and forward packets from their partners to the GW through simple amplification or decoding and forwarding~\cite{Laneman2004}. 

Instead of just acting as routers by relaying a single frame at a time, in a more elaborated cooperative mechanism the EDs may resort to the concept of network coding~\cite{Ahlswede2000} and transmit linear combinations, performed over a finite field GF($q$), of more than one frame. This approach is referred to as Network-Coded Cooperation (NCC)~\cite{Xiao2010,Rebelatto2012}, and has the purpose of improving reliability rather than the throughput, the metric usually adopted in works that resort to the network coding concept~\cite{Wu2015}.       
Against this background, in this work we propose a novel NCC-aided LoRa network, where the EDs are capable of exchanging messages with D2D communication, as well as transmitting network-coded packets to the common GW. To our knowledge, this is the first work where the concept of NCC is applied in the scope of LoRaWAN. In summary, the contributions of this paper can be listed as:
\begin{itemize}
\item We propose NCC-LoRa, a cooperative scheme where the EDs can exchange messages in a D2D fashion, as well as transmit linear combinations of different packets to the GW, resorting to the NCC concept from~\cite{Xiao2010,Rebelatto2012};
\item We analytically evaluate the outage probability of the proposed NCC-LoRa scheme, by considering both disconnection and collision events~\cite{Georgiou2017}. Moreover, differently from~\cite{Xiao2010,Rebelatto2012}, we resort to stochastic geometry tools to calculate the probability of establishing cooperation between EDs, in a multi-ring scenario where the SFs are assigned based on a target outage probability; 
\item We show, analytically and through numerical results, that NCC-LoRa can outperform both conventional LoRa (a single transmission under LoRaWAN) and LoRa with message replication (RT-LoRa) from~\cite{Hoeller2018}. More specifically, under given reliability and duty cycling requirements, we show that NCC-LoRa can simultaneously support a larger number of EDs and consume less energy than LoRa and RT-LoRa, even when taking the additional consumption of the D2D communication into account. 
\end{itemize}
The rest of this paper is organized as follows. Section~\ref{ssec:related-work} presents the state-of-the-art, by discussing some related works. Section~\ref{sec:preliminaries} introduces the system model and some relevant concepts. The outage performance of the proposed NCC-LoRa is evaluated in Section~\ref{sec:intro-ncc-lora}, while its energy consumption is studied in Section~\ref{sec:energy-efficiency}. Section~\ref{sec:numerical-results} presents some numerical results. Finally, Section~\ref{sec:final-comments} concludes the paper.

{\it Notation}: Throughout this paper, $\boxplus$ and $\boxtimes$ represent respectively a summation and a multiplication over a finite field GF($q$), where $q$ is the field size. $\mathbb{E}(x)$ is the expected value of the random variable $x$, $\mathcal{C}\mathcal{N}(a,b)$ stands for a complex normal distribution with average $a$ and variance $b$, while $\Pr\left\{\phi\right\}$ denotes the probability of event $\phi$. Also, $\lceil\cdot\rceil$ expresses the ceiling function. Finally, $|\mathcal{V}|$ represents the area of region $\mathcal{V} \in \mathbb{R}^2$. The rest of acronyms and symbols adopted in this work are summarized in Tables~\ref{tab:acronyms} and~\ref{tab:symbols}, respectively.

\begin{table}[!t]
\centering
\caption{List of acronyms adopted in this work.}
\label{tab:acronyms}
\begin{tabular}{ll}
\toprule
{\bf Acronym} & \multicolumn{1}{c}{\bf Meaning}\\
\midrule
ADR & Adaptive Data Rate\\
AWGN & Additive White Gaussian Noise\\
BER & Bit Error Rate\\
bps & Bits per second\\
CDF & Cumulative Distribution Function\\
CSS & Chirp Spread Spectrum\\
D2D & Device-to-Device\\
DR & Data Rate\\
ED & End-Device\\
FEC & Forward Error Correction\\
FSK & Frequency-Shift Keying\\
GW & GateWay\\
i.i.d. & Independent and identically distributed\\
IoT & Internet-of-Things\\
IP & Internet Protocol\\
LoRa & Long Range\\
LoRaWAN & Long Range Wide-Area Network\\
LPWAN & Low-Power Wide-Area Network\\
MDS & Maximum-Distance Separable\\
NCC & Network-Coded Cooperation\\
NS & Network Server\\
PPP & Poisson Point Process\\
RSSI & Recieved Signal Strenght Indicator\\
RT-LoRa & Long Range with Retransmissions\\
SC & Selection Combining\\
SF & Spreading Factor\\
SNR & Signal-to-Noise Ratio\\
TDMA & Time-Division Multiple Access\\
ToA & Time-on-Air\\
\bottomrule
\end{tabular}
\end{table}

\begin{table}[!t]
\centering
\caption{List of symbols adopted in this work.}
\label{tab:symbols}
\begin{tabular}{ll}
\toprule
{\bf Symbol} & \multicolumn{1}{c}{\bf Meaning}\\
\midrule
$\mathrm{S}$ & {\footnotesize Receiver sensitivity}\\
$B$ & {\footnotesize Bandwidth}\\
$\mathcal{S}_\mathrm{F}$ & {\footnotesize Spreading factor}\\
$R_\text{FEC}$ & {\footnotesize Forward error correction rate}\\
$R_s$ & {\footnotesize Transmission symbol rate}\\
$R_b$ & {\footnotesize Transmission bit rate}\\
$\text{CR}$ & {\footnotesize Coding rate}\\
$\mathrm{T}_i$ & {\footnotesize Transmission period of $i$}\\
$\text{PL}$ & {\footnotesize Payload length in bytes}\\
$n_{i}$ & {\footnotesize Number of symbols in $i$}\\
$N_b$ & {\footnotesize Number of bits}\\
$\text{DE}(.)$ & {\footnotesize Delivery optimization indicator function}\\
$\mathcal{R}$ & {\footnotesize Network range}\\
$N$ & {\footnotesize Instantaneous number of active end-devices}\\
$\bar{N}$ & {\footnotesize Average number of active end-devices}\\
$\rho$ & {\footnotesize Network density}\\
$\Phi$ & {\footnotesize Poisson Point Process}\\
$\Psi$ & {\footnotesize LoRa connection SNR threshold}\\
$d_n$ & {\footnotesize Distance from $\texttt{ED}_n$ to the gateway}\\
$\varrho$ & {\footnotesize Duty cycle}\\
$\eta$ & {\footnotesize Path loss exponent}\\
$f_c$ & {\footnotesize Carrier frequency}\\
$\lambda$ & {\footnotesize Carrier wavelength}\\
$s_n$ & {\footnotesize Signal transmitted by $\texttt{ED}_n$}\\
$p_n$ & {\footnotesize Parity packet transmitted by $\texttt{ED}_n$}\\
$P$ & {\footnotesize Transmission power}\\
$h_n$ & {\footnotesize Fading coefficient of $\texttt{ED}_n$}\\
$w$ & {\footnotesize AWGN signal}\\
$\mathcal{N}$ & {\footnotesize Noise power}\\
$\text{NF}$ & {\footnotesize Receiver noise figure}\\
$\mathcal{X}_{1,k}^{\mathcal{S}_\mathrm{F}}$ & {\footnotesize Co-SF interference indicator function}\\
$\mathcal{O}_n$ & {\footnotesize Outage probability for a single transmission from $\texttt{ED}_n$}\\
$\mathcal{O}_{\textsf{sch}}$ & {\footnotesize Outage probability for the scheme $\textsf{sch}$}\\
$\mathcal{O}_\mathrm{target}$ & {\footnotesize Network target outage probability}\\
$l_{\mathcal{S}_\mathrm{F}}$ & {\footnotesize Upper boundary of SF  $\mathcal{S}_\mathrm{F}$}\\
$\xi$ & {\footnotesize SF coverage range}\\
$\mathcal{H}_n$ & {\footnotesize Connection probability for a single transmission from $\texttt{ED}_n$}\\ 
$\mathcal{Q}_n$ & {\footnotesize Capture probability for a single transmission from $\texttt{ED}_n$}\\
$\delta$ & {\footnotesize Sum-interference capture threshold}\\
$M$ & {\footnotesize Number of transmissions within a time-slot}\\
$\mathcal{P}_\mathrm{effective}^\mathrm{D2D}$ & {\footnotesize Probability of effective D2D communication}\\[2pt]
$\mathcal{P}_\mathrm{neigh}^\mathrm{D2D}$ & {\footnotesize Neighboring probability}\\
$A_{1,\mathrm{approx}}^\mathrm{Coop}$ & {\footnotesize Cooperation area approximation}\\
$\mathcal{P}_\mathrm{1}^\mathrm{Coop}$ & {\footnotesize Probability of cooperation}\\
\bottomrule
\end{tabular}
\end{table}

\subsection{Related Works} \label{ssec:related-work}

In~\cite{Georgiou2017}, the authors resort to stochastic geometry to model the uplink coverage of a single LoRa GW in terms of two independent link outage events, namely {\it disconnection} and {\it collision}. While the former is a function of the Signal-to-Noise Ratio (SNR), the later encompasses the collision probability of EDs that simultaneously operate at the same Spreading Factor (SF), causing co-SF interference. As a result, ED coverage probability is shown to decay exponentially with the number of EDs, despite low duty cycling policy and spread spectrum orthogonality~\cite{Georgiou2017}. The model from~\cite{Georgiou2017} is then extended in~\cite{Mahmood2019} by considering also the effects of inter-SF interference in the collision probability, showing that, even though such probability increases due to the inter-SF interference, it is still dominated by the co-SF interference. Moreover, the work in~\cite{Mahmood2019} also shows that the approximated strongest interferer modelling from~\cite{Georgiou2017} may be too optimistic in dense scenarios, with several co-SF EDs.    

In~\cite{Hoeller2018}, Hoeller {\it et al.} adapted the model from~\cite{Georgiou2017} by employing message replication and multiple receive antennas at the GW aiming at achieving, respectively, time and spatial diversity. The results from~\cite{Hoeller2018} show that, while time redundancy has an optimum number of retransmissions that minimizes the overall outage probability (being in general more beneficial to low-density scenarios), the use of multiple antennas at the GW is always beneficial. In~\cite{SantAna2020}, the performance of a replication-aided LoRa network is improved even further over traditional designs by adopting the concept of coded transmission~\cite{Montejo2019}, where the EDs are capable of transmitting linear combinations of more than one of their individual frames. 

Differently from the star topology adopted in~\cite{Georgiou2017,Mahmood2019,Hoeller2018,SantAna2020,SantAna2020SIC}, in~\cite{Jiang2021} the authors propose a hybrid Mesh LPWAN, where LoRa is integrated to the short-range ANT technology~\cite{ANT} to improve the network performance in dense deployments. Communication in~\cite{Jiang2021} is performed in a multi-hop, grant-based fashion. Consequently, time division multiple access (TDMA) is required to schedule transmissions. Therefore, the design from~\cite{Jiang2021} requires extensive network-wide device synchronization, further increasing complexity and energy consumption. In~\cite{Borkotoky2019}, it is shown that the presence of relays improves the reliability of a duty cycle constrained LoRa network. In~\cite{Borkotoky2019}, the relays overhear the ED transmissions and forward them to the GW, reducing the message loss rate in up to 50\% with a single relay. However, the design in~\cite{Borkotoky2019} requires additional dedicated relays, which increases the cost of implementation.

An empirical validation of network-assisted D2D communication in LoRaWAN is carried out in~\cite{Mikhaylov2017}. With the direct exchange of messages between EDs, the authors show that time and energy consumption for data transfer can be reduced by up to 20 times when compared to regular LoRaWAN data transfer mechanisms, where such exchange is performed through the GW. In~\cite{Kim2018}, the authors address the security issue in a D2D-aided LoRaWAN by proposing key sharing between the EDs that guarantees mutual authentication, confidentiality, and integrity, while increasing the power consumption by only $5\%$ compared to the scheme from~\cite{Mikhaylov2017}.

\section{Preliminaries} \label{sec:preliminaries}
\subsection{LoRa} \label{ssec:lora}

LoRa is a spread spectrum-based modulation technique patented by Semtech~\cite{LoRaPATENT}, which aims at providing ultra-long range communication with high interference immunity while minimizing current consumption. The LoRaWAN communication protocol~\cite{LoRaWAN} is then built upon the underlying LoRa physical layer, defining the network architecture.

In networks where LoRa modulation is used, the frame Time-on-Air (ToA) and receiver sensitivity ($\mathrm{S}$) depend basically on the bandwidth $B\in\{125, 250, 500\}$ (kHz), the spreading factor (SF)  $\mathcal{S}_\mathrm{F}\in\{7,\ldots,12\}$, and forward error correction (FEC) rate $R_{\text{FEC}} = 4/(4\!+\!\text{CR})$, where $\text{CR} \in \{1,\ldots,4\}$~\cite{DatasheetLoRa2019}. The ToA, measured by the frame transmission period~$\mathrm{T}_\text{frame}(\mathcal{S}_\mathrm{F},B,R_\text{FEC})$\footnote{The transmission bandwidth $B$ and forward error correction rate $R_\text{FEC}$ are omitted in all further references for simplicity of notation.}, is then obtained as~\cite{DatasheetLoRa2019}:
\begin{equation} \label{eq:time-on-air}
\begin{split}
\mathrm{T}_\text{frame}(\mathcal{S}_\mathrm{F}) &= \mathrm{T}_{\text{pre}}(\mathcal{S}_\mathrm{F}) + \mathrm{T}_{\text{pay}}(\mathcal{S}_\mathrm{F})\\
& = (n_{\text {pre}} + 4.25)\mathrm{T}_s(\mathcal{S}_\mathrm{F}) + n_{\text {pay}}(\mathcal{S}_\mathrm{F})\mathrm{T}_s(\mathcal{S}_\mathrm{F})\\
& = \mathrm{T}_s(\mathcal{S}_\mathrm{F})\big[n_\text{pre} + n_\text{pay}(\mathcal{S}_\mathrm{F}) + 4.25\big],
\end{split}
\end{equation}
where $\mathrm{T}_{\text{pre}}(\mathcal{S}_\mathrm{F})$ and $\mathrm{T}_{\text{pay}}(\mathcal{S}_\mathrm{F})$ are the transmission periods for preamble and payload, respectively. Analogously, $n_{\text{pre}}$ and $n_{\text{pay}}(\mathcal{S}_\mathrm{F})$ correspond to the number of symbols in the preamble and payload. Also, $\mathrm{T}_s(\mathcal{S}_\mathrm{F})=2^{\mathcal{S}_\mathrm{F}}/B$ is the symbol period. According to~\cite{DatasheetLoRa2019}, $n_{\text{pay}}(\mathcal{S}_\mathrm{F})$ is given by
\begin{equation} \label{eq:n_payload}
n_{\text {pay}}(\mathcal{S}_\mathrm{F})\!=\!8+ \max\!\left\{0,\!\Biggl\lceil\!\frac{44+8\,\text{PL}\!-\!4\,\mathcal{S}_\mathrm{F}}{4\,\big(\mathcal{S}_\mathrm{F}\,-\,2\,\text{DE}(\mathcal{S}_\mathrm{F})\big)}\!\Biggr\rceil\!(4\!+\!\text{CR})\!\right\}\!,
\end{equation}
with $\text{PL}$ being the payload length (in bytes) and DE$(\mathcal{S}_\mathrm{F})$ an indicator function whose output is one for $\mathcal{S}_\mathrm{F} \geq 11$ and zero otherwise. Finally, the bit rate $R_b$ can be obtained from the symbol rate $R_s(\mathcal{S}_\mathrm{F}) = 1/\mathrm{T}_s = B/2^{\mathcal{S}_\mathrm{F}}$ as~\cite{DatasheetLoRa2019}:
\begin{equation}\label{eq:bit_rate}
R_b(\mathcal{S}_\mathrm{F}) = \mathcal{S}_\mathrm{F}\,R_{\text{FEC}}\,R_s(\mathcal{S}_\mathrm{F}) = \mathcal{S}_\mathrm{F} \, B\frac{2^{2-\mathcal{S}_\mathrm{F}}}{(4+\text{CR})}.
\end{equation}

Table~\ref{tab:lora_parameters} presents some values of ToA and receiver sensitivity, for $B = 125$ kHz and different values of SF. Note that the ToA increases exponentially with the SF, reducing the bit rate while improving the receiver sensitivity.
\begin{table}[!t]
\centering
\caption{LoRa characteristics for CR = $1$ and $B = 125$ kHz, $9$ byte payload with CRC and low data rate optimization enabled, and implicit header mode disabled~\cite{Georgiou2017}.}
\resizebox{\columnwidth}{!}{
\begin{tabular}{ccccc}
    \toprule
    {\bf SF} & {\bf Time-on-Air} & {\bf Bit Rate} & {\bf Receiver Sensitivity} & {\bf SNR Threshold}\\
        $\mathcal{S}_\mathrm{F}$ & {\footnotesize  $\mathrm{T}_\text{frame}$ (\si{\milli\second})}& {\footnotesize $R_b$ (\si{kbps})} &  $\mathrm{S}$ {\footnotesize (\si{\deci\belmilliwatt})} & $\Psi$ {\footnotesize (\si{\deci\bel})} \\
    \midrule
    $7$ & $41.22$ & $5.47$ & $-123$ & $-6$ \\
    $8$ & $72.19$& $3.13$ & $-126$ & $-9$\\
    $9$ & $144.38$ & $1.76$ & $-129$ & $-12$\\
    $10$ & $247.81$ & $0.98$ & $-132$ & $-15$\\
    $11$ & $495.62$ & $0.54$ & $-134.5$ & $-17.5$\\
    $12$ & $991.23$ & $0.29$ & $-137$ & $-20$\\
    \bottomrule
\end{tabular}
}\label{tab:lora_parameters}
\end{table}

In practice, LoRa physical layer is usually implemented along with the ALOHA-based LoRaWAN network protocol developed by the LoRa Alliance~\cite{LoRaWAN}. In LoRaWAN, the EDs communicate directly to the GWs, in a star topology. The GWs, in turn, resort to a standard Internet protocol (IP) connection to forward the decoded packets to the Network Server (NS), which is responsible for most of the tasks that require higher computational complexity. However, it is worth noting that, although being commonly used along with the LoRaWAN protocol, the use of LoRa in physical layer is in fact agnostic of higher layers.

\subsection{System Model} \label{ssec:system_model}

\begin{figure}[!t]
\centering
\includegraphics[width = 0.9\columnwidth]{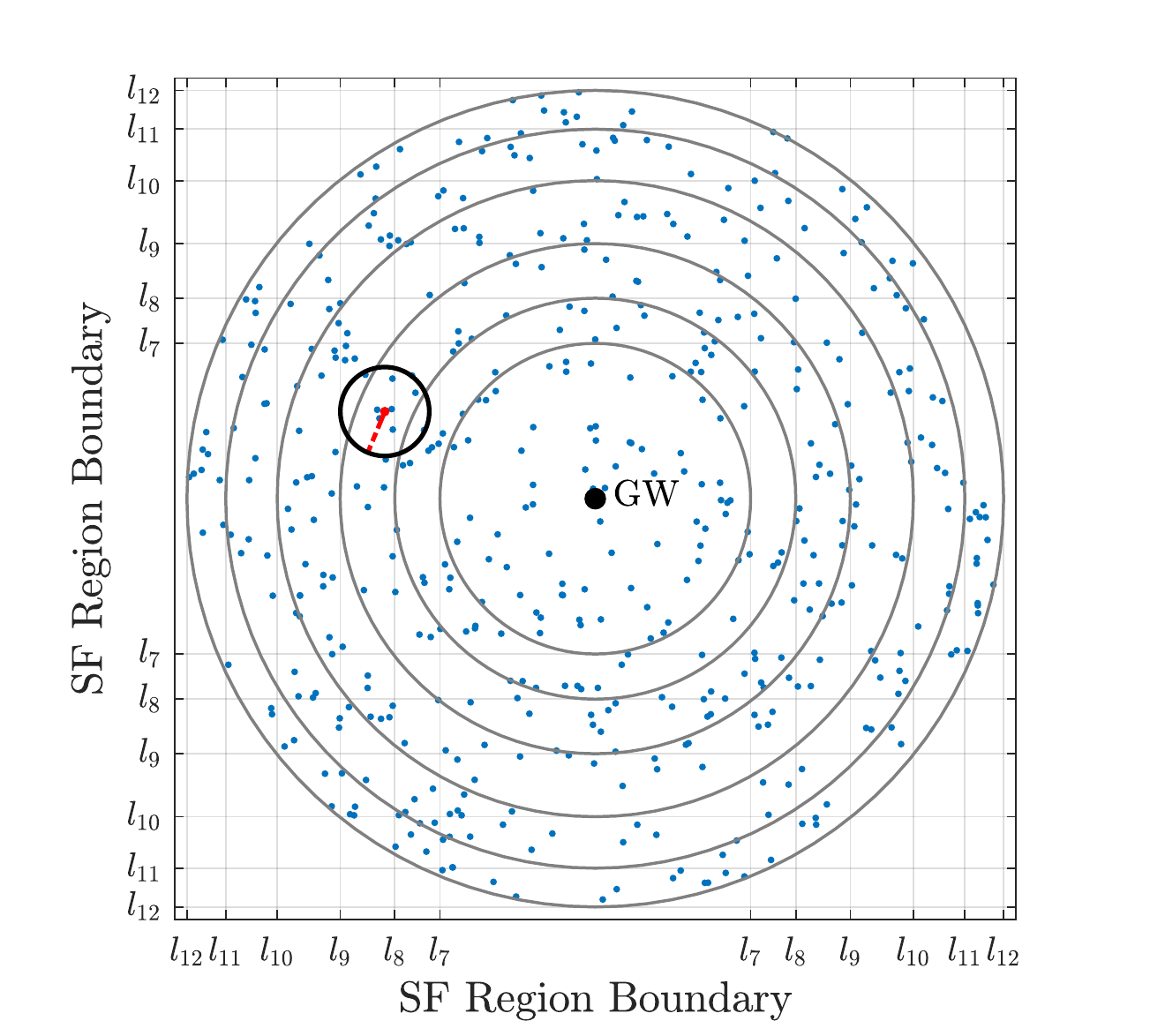}
\caption{Realization of a PPP with EDs uniformly distributed in an area of radius $\mathcal{R} = l_{12}$ around the GW. The red-dotted line illustrates the cooperation distance from $\texttt{ED}_n$, as explained in Section~\ref{ssec:d2d-phase}.}
\label{fig:system-model}
\end{figure}

We consider a scenario where a GW is placed in the center of a circular region $\mathcal{V} \in \mathbb{R}^2$ with radius $\mathcal{R}$, as illustrated in Fig.~\ref{fig:system-model}. There are also $N$ active end-devices $\{\texttt{ED}_n\}_{n=1}^{N}$ uniformly distributed in $\mathcal{V}$, according to a Poisson Point Process (PPP) $\Phi$ with intensity $\rho = \bar{N}/|\mathcal{V}|$, where $\bar{N}$ is the average number of EDs and $ |\mathcal{V}| = \pi\mathcal{R}^2$. The Euclidean distance from the randomly placed\footnote{The analysis focuses on $\texttt{ED}_1$, but conclusions apply to any ED.} $\texttt{ED}_1$ to the GW is $d_1 <\mathcal{R}$.

The uplink transmissions follow the LoRaWAN Class-A unslotted ALOHA-based grant-free channel access scheme~\cite{DatasheetLoRa2019}, with a maximum duty cycle of $1\%$ for the largest SF. Moreover, the duty cycle experienced by a given ED is 
\begin{equation}\label{eq:duty_cycle}
    \varrho(\mathcal{S}_\mathrm{F}) = \frac{\mathrm{T}_\text{frame}(\mathcal{S}_\mathrm{F})}{\mathrm{T}_{\text{slot}}},
\end{equation}
where $\mathrm{T}_\text{frame}(\mathcal{S}_\mathrm{F})$ is the ToA from~\eqref{eq:time-on-air}, which is exemplified in Table~\ref{tab:lora_parameters}, and $\mathrm{T}_{\text{slot}}$ is the duration of each time-slot. Thus, despite higher SFs requiring longer ToA, all EDs consider the same time slot duration, since we assume that they run the same application and therefore must update their measurements at the same periodicity.

The path-loss model follows the Friis equation~\cite{Goldsmith2005} 
\begin{equation} \label{eq:friis-model}
g(d_1) = \left(\frac{\lambda}{4\pi}\right)^{2}\left(d_1\right)^{-\eta},
\end{equation}
with $\lambda$ being the wavelength (\si{\meter}), $f_c$ the carrier frequency (\si{\hertz}), $d_1 > 1$~\si{\meter} the distance of $\texttt{ED}_{1}$ to the GW, and $\eta \geq 2$ the path-loss exponent.

Assuming that $\texttt{ED}_1$ transmits signal $s_1$ with power $P$ (equal to all EDs), and that the independent and identically distributed (i.i.d.) block fading is modelled as $h_1 \sim \mathcal{C}\mathcal{N}(0,1)$, \emph{i.e.}, Rayleigh fading, the signal received at the GW is~\cite{SantAna2020}
\begin{equation}\label{eq:received_signal}
r = \sqrt{P\,g(d_1)} h_1 s_1 + \sum_{k=2}^N \mathcal{X}_{1,k}^{\mathcal{S}_\mathrm{F}}\, \sqrt{P\,g(d_k)}h_k s_k + w,
\end{equation}
where $w$ is the Additive White Gaussian Noise (AWGN) with zero mean and variance $\mathcal{N} = -174 + \text{NF} + 10\log_{10}B$~(\si{\deci\belmilliwatt})~\cite{Georgiou2017}, being NF the receiver noise figure, and $\mathcal{X}_{1,k}^{\mathcal{S}_\mathrm{F}}$ is a function that indicates whether $\texttt{ED}_k$, $\forall k \!\neq\! 1$, is transmitting at the same time and frequency, and with the same SF as $\texttt{ED}_1$, thus configuring co-SF interference\footnote{Since co-SF interference causes more impact in system performance~\cite{Mahmood2019}, we do not consider inter-SF interference in this paper.}.

\subsubsection{Defining the SF Boundaries} \label{sssec:sf_allocation}
Among other alternatives present in the literature, we consider a SF allocation where the SFs are assigned following the reliability constraint $\mathcal{O}_{\textsf{sch}}(d_1, \mathcal{S}_\mathrm{F}) \leq\mathcal{O}_\mathrm{target}$, {\it i.e.}, the outage probability of $\texttt{ED}_1$, when operating under scheme $\textsf{sch} \in\{\text{LoRa, RT-LoRa, NCC-LoRa}\}$, cannot exceed the maximum allowed outage probability $\mathcal{O}_\mathrm{target}$. Thus, since $\mathcal{O}_{\textsf{sch}}(d_1,\mathcal{S}_\mathrm{F})$ increases with $d_1$ and decreases with  $\mathcal{S}_\mathrm{F}$ (see Section~\ref{ssec:outage_lora}), $\texttt{ED}_1$ uses the lowest possible SF (aiming at reducing the ToA) that meets the aforementioned reliability constraint, until reaching a boundary $l_{\mathcal{S}_\mathrm{F}} = \left\{d_1|\mathcal{O}_{\textsf{sch}}(d_1, \mathcal{S}_\mathrm{F}) =\mathcal{O}_\mathrm{target}\right\}$, as illustrated in Fig.~\ref{fig:system-model}.

Another alternatives adopted in the literature to define such boundaries include the non-optimized fixed-width approach~\cite{Raza2017,Hoeller2018,Mahmood2019}, or the received signal strength indicator (RSSI)-based approach from~\cite{Cuomo2017}, where EDs with lower RSSI adopt higher SFs. It is also worthy mentioning that the LoRaWAN protocol also includes an Adaptive Data Rate (ADR) mechanism to properly allocate transmission parameters~\cite{LoRaWAN}. However, to evaluate the performance influence of different SF allocation schemes is out the scope of this paper and is left as future work.

Finally, the coverage range $\xi(\mathcal{S}_\mathrm{F})$ of a given $\mathcal{S}_\mathrm{F}$ is defined as the difference between its lower and upper boundaries, {\it i.e.}, $\xi(\mathcal{S}_\mathrm{F}) = l_{\mathcal{S}_\mathrm{F}} - l_{\mathcal{S}_\mathrm{F}-1}$, which we also interchangeably refer to as the width of the SF.

\subsection{Uplink Outage Probability} \label{ssec:outage_lora}

The outage probability is modelled as in~\cite{Georgiou2017,Hoeller2018,SantAna2020}, such that $\texttt{ED}_1$ is in outage at the GW upon the occurrence of at least one of the following events: {\it i)} there is no connection between $\texttt{ED}_1$ and the GW; {\it ii)} $s_1$ collides with a transmission from another ED using the same SF.

Let $\mathcal{H}_1$  be the {\it connection probability} and $\mathcal{Q}_1$ be the non-collision probability, which we hereinafter refer to as {\it capture probability}. Thus, assuming that $\mathcal{H}_1$ and $\mathcal{Q}_1$ are independent, the outage probability for a single transmission from $\texttt{ED}_1$ can be approximated as~\cite{Georgiou2017,Hoeller2018,Mahmood2019,SantAna2020SIC,SantAna2020}
\begin{equation} \label{eq:outage-definition}
\mathcal{O}_1 = 1-\mathcal{H}_1\,\mathcal{Q}_1.
\end{equation}
Next we analyze $\mathcal{H}_1$ and $\mathcal{Q}_1$ from~\eqref{eq:outage-definition}, based on~\cite{Georgiou2017,Hoeller2018,Mahmood2019,SantAna2020SIC,SantAna2020}.
\subsubsection{Outage Condition 1 (Disconnection)}

The connection probability is defined as the probability that the received signal has a Signal-to-Noise Ratio (SNR) above the SNR threshold $\Psi(\mathcal{S}_\mathrm{F})$, whose values are presented in Table~\ref{tab:lora_parameters}. For Rayleigh fading, $|h_1|^2$ follows an exponential distribution, such that $\mathcal{H}_1$ becomes~\cite{Georgiou2017}: 
\begin{equation} \label{eq:connection-probability} 
\begin{split}
\mathcal{H}_1 &= \Pr\left\{\frac{P\,|h_1|^2 g(d_1)}{\mathcal{N}} \geq \Psi(\mathcal{S}_\mathrm{F})\right\} \\ 
&= \exp\left({-\frac{\mathcal{N}\Psi(\mathcal{S}_\mathrm{F})}{P\,g(d_1)}}\right).
\end{split}
\end{equation}
\subsubsection{Outage Condition 2 (Collision)}

In LoRa, it is commonly assumed that a collision only occurs if the power difference between the desired signal and any other simultaneously received signal causing co-SF interference is less than 6 \si{\deci\bel}~\cite{Georgiou2017}. However, Mahmood \textit{et. al.} showed in~\cite{Mahmood2019} that the strongest interferer model can lead to distortions when the network density $\rho$ is high. To include this conclusion in our model, we follow~\cite{Hoeller2018} and formulate the capture probability as the probability that the sum-interference caused by all collided signals is below the given $\delta = 6$~\si{\deci\bel} threshold, {\it i.e.} 
\begin{equation} \label{eq:capture-probability-definition}
\begin{split}
\mathcal{Q}_1 &= \Pr\Bigg\{\underbrace{\frac{|h_1|^2 g(d_1)}{\sum_{k=2}^{N}\mathcal{X}_{1,k}^{\mathcal{S}_\mathrm{F}}|h_{k}|^2 g(d_{k})}}_{X_k} \geq \delta \Bigg| d_1 \Bigg\}\\
& = \mathbb{E}_{|h_1|^2}\left[ \Pr\left\{\left. X_k < \frac{|h_1|^2 g(d_1)}{\delta} \right\rvert d_1 \right\}\right]\\
 &= \int_{0}^{\infty} \exp\left({-z}\right) F_{X_k}\left(\frac{z g(d_1)}{\delta}\right) \mathrm{d}z,
\end{split}
\end{equation}
where $F_{X_k}$ is the Cumulative Distribution Function (CDF) of $X_k = \sum_{k=2}^{N}\mathcal{X}_{1,k}^{\mathcal{S}_\mathrm{F}}|h_{k}|^2 g(d_{k})$. Note that~\eqref{eq:capture-probability-definition} captures the influence of collisions only. The channel and noise impairments are encompassed by the connection probability from~\eqref{eq:connection-probability}.

From~\cite{Hoeller2018}, and following the effect of unslotted-ALOHA  
on the PPP density (which can ultimately be measured by doubling the density experienced in an equivalent slotted-ALOHA model~\cite{Berioli2016,SantAna2020}), one can rewrite~\eqref{eq:capture-probability-definition} as:
\begin{equation} \label{eq:capture-probability-solved}
 \mathcal{Q}_1 =  \exp\big({-4\pi\rho\, \Lambda(d_1)\, \varrho(\mathcal{S}_\mathrm{F})}\big),
\end{equation}
where $\varrho(\mathcal{S}_\mathrm{F})$ comes from~\eqref{eq:duty_cycle}, and
\begin{equation} \label{eq:lambda_function}
\begin{split}
    \Lambda(d_1) &= \frac{\left(l_{\mathcal{S}_\mathrm{F}}\right)^2}{2}\,{}_{2}F_{1}\Big(1,\frac{2}{\eta}; 1\!+\!\frac{2}{\eta}; \frac{-\left(l_{\mathcal{S}_\mathrm{F}}\right)^{\eta}}{\delta d_1^{\eta}}\Big)\\
    &- \frac{\left(l_{\mathcal{S}_\mathrm{F}-1}\right)^2}{2}\,{}_{2}F_{1}\Big(1,\frac{2}{\eta}; 1\!+\!\frac{2}{\eta}; \frac{-\left(l_{\mathcal{S}_\mathrm{F}-1}\right)^{\eta}}{\delta d_1^{\eta}}\Big),
\end{split}
\end{equation}
with ${}_{2}F_{1}(\cdot)$ being the hypergeometric function~\cite{Daalhuis2010}, $l_{\mathcal{S}_\mathrm{F}-1}$ and $l_{\mathcal{S}_\mathrm{F}}$ the inner and outer radii of the SF region where $\texttt{ED}_1$ is situated, respectively. Finally, after replacing~\eqref{eq:connection-probability} and~\eqref{eq:capture-probability-solved} in~\eqref{eq:outage-definition}, the outage probability becomes 
\begin{equation} \label{eq:outage-lora}
\begin{split}
\mathcal{O}_1 & = 1-\mathcal{H}_1\mathcal{Q}_1\\
& = 1 - \exp\left({-4\pi\rho \,\Lambda(d_1) \,\varrho(\mathcal{S}_\mathrm{F})- \frac{\mathcal{N}\Psi(\mathcal{S}_\mathrm{F})}{P\,g(d_1)}}\right)\\
& \approx 4\pi\rho\,\Lambda(d_1)\,\varrho(\mathcal{S}_\mathrm{F}) + \frac{\mathcal{N}\Psi(\mathcal{S}_\mathrm{F})}{P\,g(d_1)},
\end{split}
\end{equation}
where $1-\exp(-x) \approx x$ holds for small values of $x$. 

\subsection{LoRa with Retransmissions (RT-LoRa)} \label{ssec:rt-lora}

We consider as benchmark the RT-LoRa design~\cite{Hoeller2018,SantAna2020}, a replication-aided scheme where each ED transmits their frames $M\geq 1$ times. Upon receiving $M$ copies of the same message, we consider that the GW applies Selection Combining (SC)~\cite{Goldsmith2005}, whose outage probability is~\cite{Hoeller2018,SantAna2020}
\begin{equation}\label{eq:outage-rt-lora}
    \mathcal{O}_\mathrm{RT-LoRa}= \left(\mathcal{O}_{1}\right)^M = \Big(1 - \mathcal{H}_1\mathcal{Q}_1\Big)^M,
\end{equation}
where $\mathcal{O}_{1}$ comes from~\eqref{eq:outage-lora}. It is worthy mentioning that, since RT-LoRa requires $M$-times as many transmissions as LoRa, one needs to properly adjust the duty cycling policy from~\eqref{eq:duty_cycle} in order not to exceed the maximum value of $1\%$. Moreover, we consider the scenario where the $M$ transmissions occur within a single time-slot\footnote{Although hybrid-coded transmission schemes~\cite{SantAna2020} can yield better performance than RT-LoRa, they require the use of multiple time-slots. In our analysis, we assume that data is continuously generated (there is new information to transmit to the GW at every time-slot) and it is time-sensitive on the $T_\mathrm{slot}$ scale. Then, the use of multiple time-slots becomes impractical.}, but are separated in time so that the channel coherence time is respected, resulting in independent channel realizations for each transmission~\cite{SantAna2020}. 

\subsection{Network-Coded Cooperation (NCC)} \label{ssec:ncc}

In an uplink NCC-aided network, nodes exploit the broadcast nature of the wireless channel to mutually help each other by relaying information. The data to be relayed by a given node is a linear combination performed over a non-binary finite field GF($q$) of its own information and the information from other cooperating partners. This process occurs in all cooperating nodes independently~\cite{Xiao2010,Rebelatto2012}.

In the NCC scheme from~\cite{Xiao2010,Rebelatto2012}, the transmission process is divided in two phases: {\it i)} the broadcast phase, where the nodes broadcast their individual information; and {\it ii)} the cooperative phase, where the nodes transmit parity frames composed of independent linear combinations over GF($q$) of their original data with those from their cooperating partners.

Let us consider a scenario where two cooperating nodes, say $\texttt{ED}_1$ and $\texttt{ED}_2$, have independent messages $s_1$ and $s_2$ of length $N_b$ bits to transmit to a common destination.  By considering a network code with rate $R_{\text{NCC}} = 1/2$, where node $m \in\{1,2\}$ transmits $s_m$ in the broadcast phase followed by the transmission of one parity frame $p_m$ in the cooperation phase, the set $\textbf{r} = [s_1^T \quad s_2^T \quad p_1^T \quad p_2^T]$ of messages potentially received at the destination is~\cite{Rebelatto2012}:
 \begin{equation} \label{eq:example-ncc}
    \begin{split}
        \textbf{r} &=  \left[
          \begin{array}{c}
        s_1\\ s_2\\
          \end{array}
        \right]^{T}
        \left[
          \begin{array}{cccc}
            1 & 0 & 1 & 1 \\
            0 & 1 & 1 & 2 \\
          \end{array}
        \right]
          =   \left[
          \begin{array}{l}
            s_1 \\ 
            s_2 \\ 
            s_1\boxplus s_2 \\ 
            s_1\boxplus 2 \boxtimes s_2 \\
          \end{array}
        \right]^T,
    \end{split}
\end{equation}
where $\boxplus$ and $\boxtimes$ represent respectively a summation and a multiplication over a finite field GF($4$), \textit{i.e.}, $p_2 = s_1 \boxplus 2 \boxtimes s_2$ represents the addition over GF($4$) of $s_1$ and $s_2$ multiplied by the linear coefficient $2 \in \text{GF}(4)$, following, for example, the set of results presented in Tables~\ref{tab:summation_GF4} and~\ref{tab:multiplication_GF4}.  Let us consider, for illustration purposes, a simplified scenario with $N_b = 8$ where $s_1$ and $s_2$ are given respectively by
 \begin{align*} 
s_1 &=  [\, 1 \enskip 1 \enskip 0 \enskip 1 \enskip 1 \enskip 0 \enskip 0 \enskip 1 \,]_{2} = [\, 3 \enskip 1 \enskip 2 \enskip 1 \,]_{4}    \\ 
s_2 &=  [\, 0 \enskip 1 \enskip 0 \enskip 0 \enskip 0 \enskip 1 \enskip 1 \enskip 1 \,]_{2} = [\, 1 \enskip 0 \enskip 1 \enskip 3 \,]_{4},
\end{align*}
where $[\cdot]_{q}$ represents the message in GF($q$). Thus, following the addition and multiplication results presented in Table~\ref{tab:GF4}, one has that the network coding operation is performed in a symbol-by-symbol basis as detailed below
 \begin{equation*} 
 \begin{split}
s_1 \boxplus s_2 &= [\, 3\boxplus1 \quad 1\boxplus0 \quad 2\boxplus1 \quad 1\boxplus3 \,]_{4}\\
&=  [\, 2 \enskip 1 \enskip 3 \enskip 2 \,]_{4} \\
&= [\, 1 \enskip 0 \enskip 0 \enskip 1 \enskip 1 \enskip 1 \enskip 1 \enskip 0 \,]_{2}       
 \end{split}
\end{equation*}
and
 \begin{equation*}
 \begin{split}
s_1 \boxplus 2\boxtimes s_2 &= [\, 3\boxplus2\boxtimes1 \quad 1\boxplus2\boxtimes0 \quad 2\boxplus2\boxtimes1 \quad 1\boxplus2\boxtimes3 \,]_{4}\\
&= [\, 3\boxplus2 \quad 1\boxplus0 \quad 2\boxplus2 \quad 1\boxplus1 \,]_{4}\\
&= [\, 1 \enskip 1 \enskip 0 \enskip 0 \,]_{4}\\
&= [\, 0 \enskip 1 \enskip 0 \enskip 1 \enskip 0 \enskip 0 \enskip 0 \enskip 0 \,]_{2}. \\
\end{split}
\end{equation*}

\begin{table}[!t]
     \caption{Operations in the finite field GF($4$).}
     \label{tab:GF4}
\hfill
    \begin{subtable}[h]{0.23\textwidth}
\centering
\begin{tabular}{c|cccc}
    \toprule
    $\boxplus$ & $0$ & $1$ & $2$ & $3$\\
    \midrule
    $0$ & $0$ & $1$ & $2$ & $3$\\
    $1$ & $1$ & $0$ & $3$ & $2$\\
    $2$ & $2$ & $3$ & $0$ & $1$\\
    $3$ & $3$ & $2$ & $1$ & $0$\\
    \bottomrule
\end{tabular}%
\vspace{1.5ex}
\caption{Addition.}
\label{tab:summation_GF4}
    \end{subtable}
    \hfill
    \begin{subtable}[h]{0.23\textwidth}
\centering
\begin{tabular}{c|cccc}
    \toprule
    $\boxtimes$ & $0$ & $1$ & $2$ & $3$\\
    \midrule
    $0$ & $0$ & $0$ & $0$ & $0$\\
    $1$ & $0$ & $1$ & $2$ & $3$\\
    $2$ & $0$ & $2$ & $3$ & $1$\\
    $3$ & $0$ & $3$ & $1$ & $2$\\
    \bottomrule
\end{tabular}
\vspace{1.5ex}
\caption{Multiplication.}
\label{tab:multiplication_GF4}
     \end{subtable}
     \hfill
     \vspace{-1em}
\end{table}

As discussed in~\cite{Rebelatto2012}, the coefficients of the linear combinations, belonging to the set $\{0,\ldots,q\!-\!1\}$, are chosen in order to form a maximum-distance separable (MDS) code, and consequently guarantee maximum achievable diversity order.

Upon receiving the messages from the set $\textbf{r}$, the destination is capable of recovering $s_1$ and $s_2$ from any two out of the four received frames. Thus, $s_1$ will be in outage at the destination when the direct transmission of $s_1$  is not decoded, and when at least two of the remaining three messages are in outage. This event occurs with probability~\cite{Rebelatto2012}: 
\begin{equation}\label{eq:outage-ncc}
\begin{split}
\mathcal{O}_\mathrm{NCC} &= \mathcal{O}_1\Big(\mathcal{O}_1\mathcal{O}_2^2 + 2\mathcal{O}_1\mathcal{O}_2(1-\mathcal{O}_2) + \mathcal{O}_2^2(1-\mathcal{O}_1)\Big) \\
&= 2\mathcal{O}_1^2\mathcal{O}_2 + \mathcal{O}_1\mathcal{O}_2^2 - 2\mathcal{O}_1^2\mathcal{O}_2^2,
\end{split}
\end{equation}
where $\mathcal{O}_\text{1}$ and $\mathcal{O}_\text{2}$ are the outage probability of a single transmission from $\texttt{ED}_1$ and the cooperating partner $\texttt{ED}_2$, respectively. The outage probability of $s_2$ is obtained similarly to~\eqref{eq:outage-ncc} by swapping sub-indices $_1$ and $_2$.

When both EDs are subjected to the same average SNR ({\it e.g.} when $d_1 \approx d_2$), one has that~\eqref{eq:outage-ncc} can be approximated at the high-SNR regime as  
\begin{equation}\label{eq:outage-ncc-app}
\mathcal{O}_\mathrm{NCC} \approx 3\mathcal{O}_1^3,
\end{equation}
indicating that NCC can achieve a higher diversity order than the one achieved by RT-LoRa (NCC achieves diversity 3 instead of diversity 2 achieved by RT-LoRa with $M=2$).

\section{Proposed NCC-LoRa Scheme} \label{sec:intro-ncc-lora}

\begin{figure*}
     \centering
     \begin{subfigure}[!t]{0.3\textwidth}
         \centering
         \includegraphics[width=\textwidth]{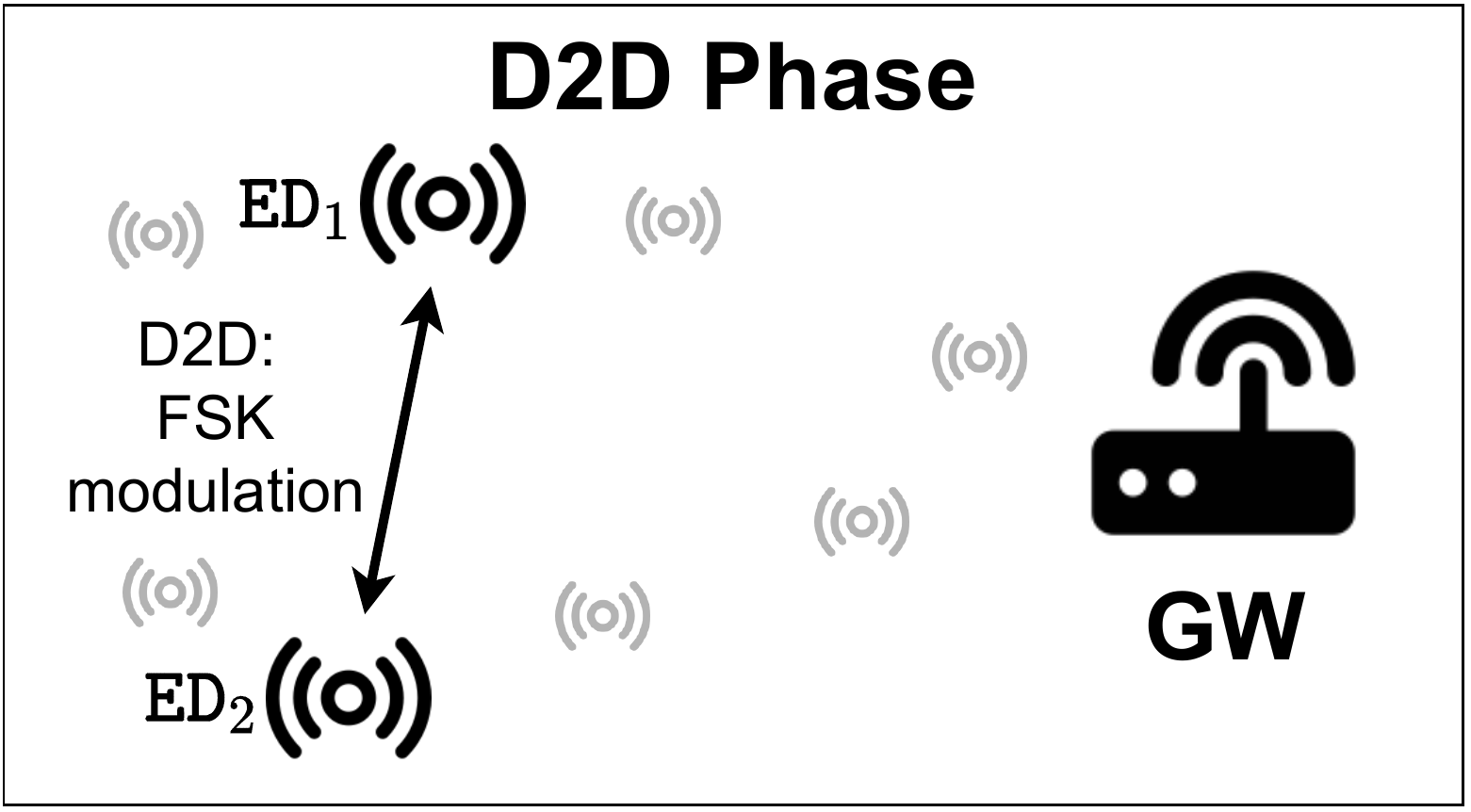}
         \caption{}
         \label{fig:ncc-lora-model-a}
     \end{subfigure}
     \hfill
     \begin{subfigure}[!t]{0.3\textwidth}
         \centering
         \includegraphics[width=\textwidth]{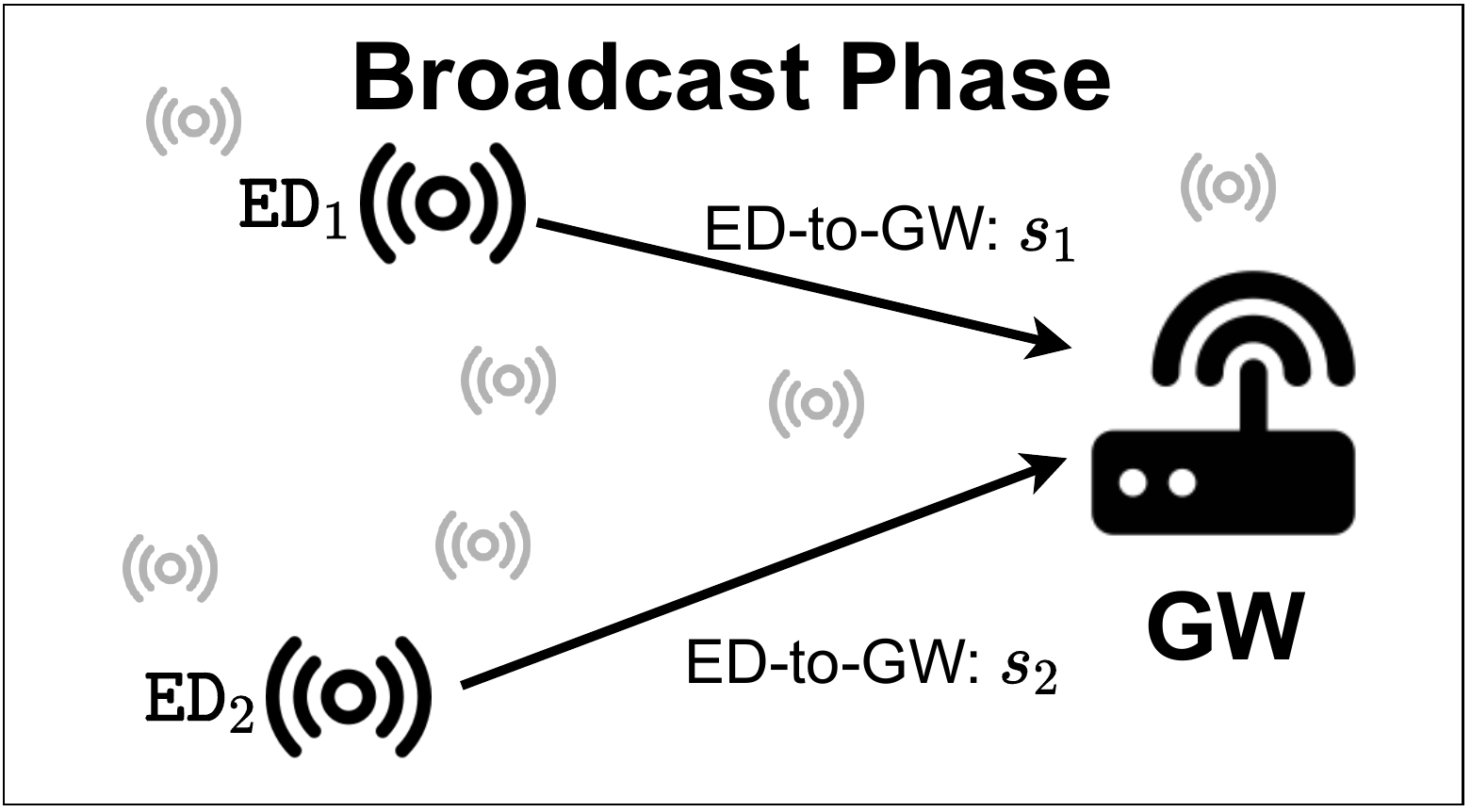}
         \caption{}
         \label{fig:ncc-lora-model-b}
     \end{subfigure}
     \hfill
     \begin{subfigure}[!t]{0.3\textwidth}
         \centering
         \includegraphics[width=\textwidth]{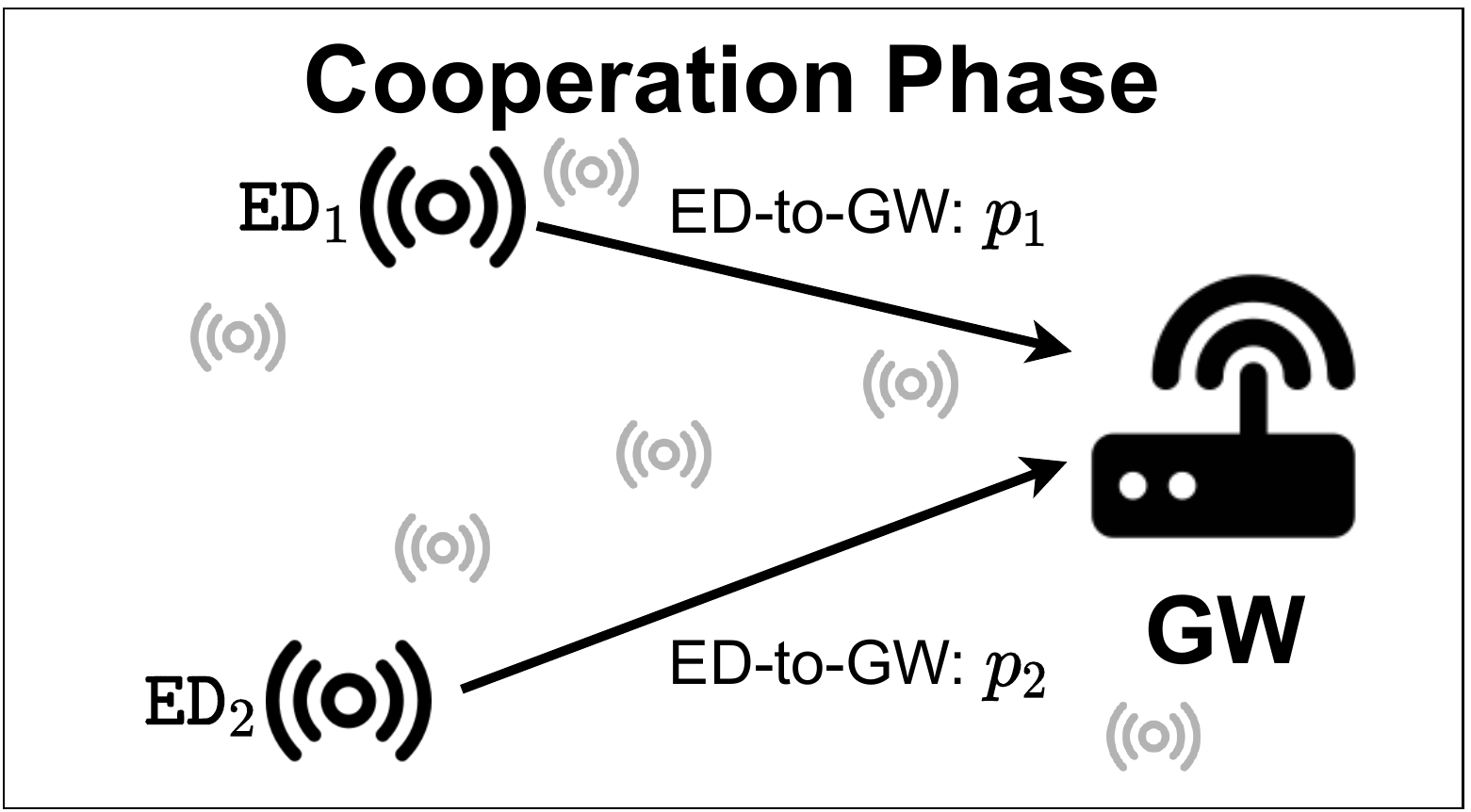}
         \caption{}
         \label{fig:ncc-lora-model-c}
     \end{subfigure}
        \caption{NCC-LoRa protocol: (a) EDs exchange their packets via D2D communication adopting a high-rate FSK modulation, with limited coverage and low ToA; (b) EDs transmit their information packets (say, $s_1$ and $s_2$) to the GW, adopting the proper SF depending on their distance from the GW; (c) $\texttt{ED}_1$ and $\texttt{ED}_2$ transmit respectively the parity packets $p_1$ and $p_2$, which are linear combinations of the information packets previously transmitted in the broadcast phase.}
        \label{fig:ncc-lora-model}
\end{figure*}

\subsection{D2D Communication} \label{ssec:d2d-phase}

The establishment of NCC requires the exchange of information frames between cooperating EDs, as illustrated in Fig.~\ref{fig:ncc-lora-model-a}. In our work, we consider that the D2D communication is performed using frequency-shift keying (FSK) modulation, mainly due to the following reasons:  {\it i)} First of all, FSK is by default available in the Semtech SX1272 transceiver, without requiring any additional hardware; {\it ii)} The FSK mode in the Semtech SX1272 transceiver allows higher data rates (hundreds of kbps) than LoRa, reducing the ToA and, consequently, leading to a marginal increase in the overall energy consumption of the network; and {\it iii)} Its interference in the LoRa communication adopted in the main link between EDs and GW is lower when adopting FSK than when adopting LoRa modulation, mainly also due to the reduced ToA. As a drawback of using FSK in the D2D communication one could mention its reduced communication range in relation to LoRa.

The probability of successful D2D communication, which we refer to as \textit{cooperation probability}, can be measured by taking into consideration\footnote{D2D neighbor discovery and message exchange are out of the scope of this work. However, in practice, an initial access protocol for finding nearby EDs, as used in Bluetooth Low Energy~\cite{BLE_NDP}, could be implemented.}: $i)$ the success probability of data transferring and $ii)$ the \textit{neighboring probability}, or the probability that EDs find potential cooperation partners in their proximity, within the D2D range.

\subsubsection{D2D link outage probability} 
The probability of correct data transferring in the D2D link is defined as:
\begin{equation}\label{eq:outage-d2d}
    \mathcal{P}_\mathrm{effective}^\mathrm{D2D} = 1- \mathcal{O}_\mathrm{D2D},
\end{equation}
where $\mathcal{O}_\mathrm{D2D}$ is the outage probability of the D2D phase.

\subsubsection{Neighboring probability} 
The probability that an ED is able encounter a neighboring ED to cooperate, $\mathcal{P}_\mathrm{neigh}^\mathrm{D2D}$, depends basically on the \textit{cooperation distance} and \textit{cooperation area}, which are defined in what follows.

\begin{definition}{Cooperation Distance.} \label{def:cooperation-distance}
The cooperation distance of $\texttt{ED}_1$ is the distance at which its transmissions can be received with an average power that is equal to the receiver sensitivity, $d_1^{\mathrm{Coop}}: 10\log_{10}g(d_1^{\mathrm{Coop}}) = \mathrm{S}_\text{D2D} (\text{\si{\deci\belmilliwatt}}) - P_\text{D2D} (\text{\si{\deci\belmilliwatt}})$.
\end{definition}

From the Friis path-loss model in~\eqref{eq:friis-model}, $d_1^\mathrm{Coop}$ becomes: 
\begin{equation} \label{eq:cooperation-distance}
d_1^{\mathrm{Coop}} = \left(\frac{\lambda}{4\pi}\right)^{\frac{2}{\eta}}10^{-\frac{ S_\text{D2D} - P_\text{D2D}}{10\eta}},
\end{equation}
which, for $\mathrm{S}_\text{D2D} = -82$~\si{\deci\belmilliwatt}, $P_\text{D2D} = 13$~\si{\deci\belmilliwatt}~\cite{DatasheetLoRa2019},  $f_c = 868$~\si{\mega\hertz} and $\eta = 2.7$~\cite{Georgiou2017}, $d_1^{\mathrm{Coop}} \approx 230$~\si{\meter}\footnote{Note that $d_1^{\mathrm{Coop}}$ represents the maximum distance beyond which cooperation cannot be established. It does not mean that the cooperative EDs are necessarily separated by such distance. Moreover, note also that, due to the fading randomness, cooperation can be beneficial even when the D2D distance is larger than the distance between the typical ED to the GW~\cite{Laneman2004}.}. This is illustrated by the red-dotted line in Fig.~\ref{fig:system-model}. 

In our model, for the sake of simplicity, we assume that D2D communication is established only between EDs adopting the same SF. Thus, the cooperation area is defined as given below. 
\begin{definition}{Cooperation Area.} \label{def:cooperation-area}
The cooperation area $A_1^{\mathrm{Coop}}$ is the area of the intersection between the circle of radius $d_1^{\mathrm{Coop}}$ centered at $\texttt{ED}_1$ and the annular SF region where $\texttt{ED}_1$ is located.
\end{definition}

The cooperation area is limited by both the cooperation distance and the SF region limits. The exact analytical evaluation of the cooperation area is complex due to high number of possible interactions between the intersected regions. Next we present an approximated value to the cooperation area.
\begin{lemma}\label{lem:cooperation-area-approx}
    The cooperation area can be approximated by
    \begin{equation}\label{eq:Ac-approx}
    A_{1,\mathrm{approx}}^\mathrm{Coop} = \min \left\{\frac{\pi}{2}(d_1^{\mathrm{Coop}})^2,2 d_1^{\mathrm{Coop}} \xi(\mathcal{S}_{\mathrm{F}}) \right\}.
    \end{equation}
\end{lemma}
\begin{IEEEproof}
Please refer to Appendix~\ref{ap:cooperation-area-approx}.
\end{IEEEproof}

Thus, the neighboring probability is formally defined as
\begin{definition}{Neighboring Probability.} \label{def:neighboring-probability}
The neighboring probability of $\texttt{ED}_1$ is as $\mathcal{P}_\mathrm{neigh}^{\mathrm{D2D}} = \Pr\{N_\mathrm{neigh}^\mathrm{Coop} \geq 1\}$, where $N_\mathrm{neigh}^\mathrm{Coop}$ is the number of EDs within the cooperation area of $\texttt{ED}_1$. 
\end{definition}

\begin{lemma} \label{lem:neighboring-probability}
 The neighboring probability of $\texttt{ED}_1$ is
 \begin{equation} \label{eq:neighboring-probability}
 \mathcal{P}_\mathrm{neigh}^{\mathrm{D2D}} = 1 - \exp\left(-\rho\, A_1^{\mathrm{Coop}}\right).
 \end{equation}
\end{lemma}
\begin{IEEEproof}
The proof follows from the nearest neighbor analysis of a stochastic PPP~\cite{Daley2013}.
\end{IEEEproof}

Finally, the cooperation probability is obtained with the aid of~\eqref{eq:outage-d2d} and~\eqref{eq:neighboring-probability} as\footnote{One should note that, even though the cooperation distance from Definition~\ref{def:cooperation-distance} depends only on the FSK modulation adopted in the D2D communication, the cooperation probability turns out to depend on the LoRa communication adopted in the link between EDs to GW due to the assumption that the EDs are able to cooperate only with EDs adopting the same SF.}
\begin{equation}\label{eq:cooperation-probability}
\begin{split}
        \mathcal{P}_1^\mathrm{Coop} &= \mathcal{P}_\mathrm{effective}^\mathrm{D2D}\,\mathcal{P}_\mathrm{neigh}^\mathrm{D2D}\\
        &= \big(1-\mathcal{O}_\mathrm{D2D}\big)\left[1 - \exp\left(-\rho A_1^\mathrm{Coop}\right)\right].
\end{split}
\end{equation}
\subsection{EDs-to-GW Communication} \label{ssec:ncc-lora}

In the proposed method, after the successful establishment of D2D communication with a neighbor, the EDs resort to NCC to improve the link quality with respect to the GW, following the approach presented in Section~\ref{ssec:ncc} and depicted in Fig.~\ref{fig:ncc-lora-model}. In case D2D communication cannot be established, we consider that the EDs switch to the RT-LoRa method presented in Section~\ref{ssec:rt-lora}\footnote{Note that this switch from NCC-LoRa to RT-LoRa can be easily implemented in practice by replacing the linear coefficients related to the cooperating partner by zero when composing the parity packet.}. Thus, regardless whether transmitting a linear combination or a replication of its own message, each ED transmits directly to the GW $M=2$ times per time slot, in the so-called broadcast phase (Fig.~\ref{fig:ncc-lora-model-b}) and cooperation phase (Fig.~\ref{fig:ncc-lora-model-c}). Hence, NCC-LoRa does not configure a multi-hop scheme as defined by~\cite{Broch1998},  but instead a single-hop scheme towards the GW aided by D2D communication between a pair of EDs.

\begin{theorem} \label{th:outage_ncc_lora}
The outage probability of NCC-LoRa is 
\begin{equation}\label{eq:outage-ncc-lora}
\begin{split}
\mathcal{O}_\mathrm{NCC-LoRa} = \mathcal{O}_{1}^{2} - \mathcal{P}_{1}^\mathrm{Coop}&\big(\mathcal{O}_{1}^{2} + 2\mathcal{O}_{1}^{2}\mathcal{O}_{2}^{2} \\ &-2\mathcal{O}_{1}^{2}\mathcal{O}_{2} - \mathcal{O}_{1}\mathcal{O}_{2}^{2}\big),
\end{split}
\end{equation} 
where $\mathcal{O}_1$ and $\mathcal{O}_2$ are the outage probabilities of a single transmission from~\eqref{eq:outage-lora} for $\texttt{ED}_1$ and its cooperating partner $\texttt{ED}_2$, respectively, and $\mathcal{P}_1^\mathrm{Coop}$ is the cooperation probability from~\eqref{eq:cooperation-probability}. 
\end{theorem}
\begin{IEEEproof}
When an ED does not establish a D2D link, its operation follows the RT-LoRa scheme. Thus, the NCC-LoRa outage probability in~\eqref{eq:outage-ncc-lora} is derived from the independent link outage probabilities of NCC and RT-LoRa by applying the complement rule to the probability of cooperation:
\begin{equation}\label{eq:outage-ncc-lora-definition}
\mathcal{O}_{\mathrm{NCC-LoRa}} = \mathcal{P}_1^{\mathrm{Coop}}\mathcal{O}_{\mathrm{NCC}} + \big(1-\mathcal{P}_1^{\mathrm{Coop}}\big)\mathcal{O}_{\mathrm{RT-LoRa}}.
\end{equation} 
\end{IEEEproof}

One can see from~\eqref{eq:outage-ncc-lora-definition} that the outage probability of the proposed NCC-LoRa tends to the outage probability of RT-LoRa when cooperation cannot be established\footnote{Another different but mathematical identical scenario is the {\it decode-and-forward} (DF) scheme~\cite{Laneman2004}, where cooperation is established but the devices forward to the common destination only the information of their partners, without resorting to network coding. However, since DF requires D2D communication while achieving the same outage performance of RT-LoRa, it is not considered among the benchmark schemes in this work, as it would lead to an increased power consumption.}. Regarding the implementation of NCC-LoRa, it is worthy remarking that:

\begin{remark} \label{rem:ncc_blocks}
After each communication round ({\it i.e.}, BP + CP phases), the GW first attempts to decode each of the four packets ($s_1$, $s_2$, $s_1 \boxplus s_2$ and $s_1 \boxplus 2 \boxtimes s_2$) individually, following the regular LoRa procedure, after properly despreding the signals. The decoded packets are then forwarded to the Network Server (NS). In a next step, the NS can resort, {\it e.g.}, to Gaussian elimination to recover both information frames $s_1$ and $s_2$ out of any two out of the four received frames. Thus, besides the additional network coding/decoding modules required by NCC-LoRa, all the remaining steps in the transmission process are in accordance to the usual LoRa guidelines ({\it e.g.} spread spectrum, FEC, etc.).  
\end{remark}
\begin{remark} \label{rem:ncc_overhead}
NCC-LoRa requires a negligible additional overhead: the typical ED only needs to append a single additional bit to inform the GW whether a given frame corresponds to a network-coded frame (parity frame) or an information frame (in case cooperation could not be established, for instance), and a few more bits to indicate the partner ED it is cooperating with. Note that the EDs do not necessarily need to transmit the coefficients adopted to generate the linear combinations, since such coefficients are generally fixed and can be previously defined and stored at the GW~\cite{Rebelatto2012}. Moreover, it is worthy mentioning that NCC-LoRa requires a larger buffer to store two frames at a time, as opposite to a single frame in traditional LoRaWAN.
\end{remark}

Due to the much lower ToA of FSK when compared to the $M=2$ LoRa transmissions of NCC-LoRa performed in the ED-to-GW link (less than 0.6\% for $\mathcal{S}_\mathrm{F}=7$ and approximately 0.02\% for $\mathcal{S}_\mathrm{F}=12$), we neglected the influence of the D2D communication in the duty cycle of the NCC-LoRa. Moreover, it is worthy mentioning that, for fair comparison purposes, besides limiting the maximum duty cycle to 1\%, the additional transmissions in the ED-to-GW link performed by RT-LoRa and NCC-LoRa must be computed into the PPP $\Phi$, as they contribute to the channel load~\cite{Hoeller2018,SantAna2020}. Thus, the average number of active transmissions is $M\bar{N}$, increasing the collision probability. 
\section{Energy Consumption of NCC-LoRa}\label{sec:energy-efficiency}

In a dense network the high probability of cooperation will push EDs into using the NCC mode. Although this may lead to a reduced outage probability, an increased energy consumption can also be expected, as more EDs perform the additional D2D transmissions. Due to the relevance of energy consumption in LoRaWAN, it becomes  fundamental to analyze this aspect. In what follows, we explore this energy-outage trade-off in more detail by modelling the average current consumption for NCC-LoRa, RT-LoRa and conventional LoRa (which corresponds to the conventional way that LoRa technology is used by the LoRaWAN network protocol, and whose outage probability comes from~\eqref{eq:outage-lora}) schemes. 
\subsection{Energy Consumption Model}

\begin{table}[!t]
\caption{LoRa consumption~\cite{DatasheetLoRa2019,LoRaCalculator}.}
\resizebox{\columnwidth}{!}{%
\begin{tabular}{@{}clcc@{}}
\toprule
\textbf{State ID} &
\multirow{2}{*}{\textbf{Description}} &
\textbf{Duration} &
\textbf{Current Consumption}\\ 
{\footnotesize $i$} &
&
{\footnotesize $\mathrm{T}_i$} &
{\footnotesize $I_i$}\\ 
\midrule
1  & Sleep                          & Eq.~\eqref{eq:sleep-period}                               & $100$~\si{\nano\ampere}               \\
2  & Standby                        & $250$~\si{\micro\second}                                  & $1.5$~\si{\milli\ampere}              \\
3  & TX frequency synthesis            & $60$~\si{\micro\second}                                   & $4.5$~\si{\milli\ampere}              \\
\multirow{2}{*}{4}  & \multirow{2}{*}{Transmission}                   & \multirow{2}{*}{Eq.~\eqref{eq:time-on-air}}                               & $22$~\si{\milli\ampere}   \scriptsize{($P= 0$ \si{\deci\belmilliwatt})} \\
  &                   &                                & $32$~\si{\milli\ampere} \scriptsize{($P=11$ \si{\deci\belmilliwatt})} \\
\bottomrule
\end{tabular}
}\label{tab:lora-energy-consumption}
\end{table}

\begin{table}[!t]
\caption{FSK consumption for $P=13$~\si{\deci\belmilliwatt}, $N_b = 120$ bits, $R_\text{D2D}=250$ kbps, $B_\text{D2D} = 250~\si{\kilo\hertz}$ and a 10~\si{\micro\second} transmission mode ramp up time~\cite{DatasheetLoRa2019}.}
\resizebox{\columnwidth}{!}{%
\begin{tabular}{@{}clcc@{}}
\toprule
\textbf{State ID} &
\multirow{2}{*}{\textbf{Description}} &
\textbf{Duration} &
\textbf{Current Consumption}\\ 
{\footnotesize $i$} &
&
{\footnotesize $\mathrm{T}_i^\text{FSK}$} &
{\footnotesize $I_i^\text{FSK}$}\\ 
\midrule
1   & Sleep                         & $\mathrm{T}_\text{slot} - \sum_{i=2}^6 \mathrm{T}_i^\text{FSK}$                     & $100$~\si{\nano\ampere}       \\      
2   & Standby                       & $250$~\si{\micro\second}              & $1.5$~\si{\milli\ampere}      \\
3   & TX frequency synthesis        & $60$~\si{\micro\second}               & $4.5$~\si{\milli\ampere}      \\
4   & Transmit                      & $499.5$~\si{\micro\second}              & $28$~\si{\milli\ampere}       \\
5   & TX to RX turnaround           & $50$~\si{\micro\second}               & $4.5$~\si{\milli\ampere}      \\
6   & Receive                       & $543$~\si{\micro\second}              & $11.2$~\si{\milli\ampere}     \\
\bottomrule
\end{tabular}%
}\label{tab:fsk-energy-consumption}
\end{table}

We consider the values for current and duration of each operational state for a LoRa transmission as measured in~\cite[Tables~6, 7 and 10]{DatasheetLoRa2019}, whose relevant values are reproduced in Table~\ref{tab:lora-energy-consumption}, which refers to a Semtech SX1272 transceiver. Furthermore, we model the consumption of both RT-LoRa and NCC-LoRa schemes so that, in order to improve efficiency, the ED performs $M$ transmissions to the GW consecutively within the same time-slot, where $M=1$ for conventional LoRa and $M=2$ for RT-LoRa and NCC-LoRa, as in~\cite{SantAna2020}. Consequently, the average current in a transmission cycle is~\cite{SantAna2020}: 
\begin{equation}\label{eq:time-slot-average-current}
    I_\mathrm{avg}(M) = \frac{1}{\mathrm{T}_{\text{slot}}}\Biggl[M\mathrm{T}_4 I_4+{\sum_{i=1}^{3}}\,\mathrm{T}_i I_i\Biggr],
\end{equation}
where each state duration $\mathrm{T}_i$ and corresponding current consumption $I_i$ values are in Table~\ref{tab:lora-energy-consumption}. Also, the duration of the sleep state is  calculated as~\cite{SantAna2020}: 
\begin{equation}\label{eq:sleep-period}
    \mathrm{T}_{1} =\,\mathrm{T}_\text{slot} - M\mathrm{T}_4 - {\sum_{i=1}^{3}}\,\mathrm{T}_i.
\end{equation}

The average current consumption of each scheme is then given in what follows. 
\subsubsection{Conventional LoRa}
In conventional LoRa, there is a single transmission per time-slot\footnote{Recall that all schemes share a common time-slot duration. This reduces the effective duty cycle for the conventional LoRa model by a factor of $1/2$ when compared to the proposed NCC-LoRa.}. Therefore, its average consumption is obtained by setting $M=1$ in~\eqref{eq:time-slot-average-current} and~\eqref{eq:sleep-period}.

\subsubsection{RT-LoRa}
Since RT-LoRa makes $M$ transmissions per time-slot, the average current is~\eqref{eq:time-slot-average-current}
\begin{equation}\label{eq:rt-lora-average-current}
    I_\mathrm{avg}^\mathrm{RT-LoRa} = I_\mathrm{avg}(M).
\end{equation}


\begin{table}[!t]
\centering
\caption{System parameters.}
\resizebox{\columnwidth}{!}{%
\begin{tabular}{llll}
    \toprule
    {\bf Parameter} & {\bf Variable} & {\bf Value } & {\bf Reference}\\    
    \midrule
    Maximum duty cycle                  & & 1$\%$ &~\cite{SantAna2020}\\
    Coding rate                         & CR & 1 &~\cite{Georgiou2017}\\
    Payload length                      & PL & 9~bytes &~\cite{Hoeller2018}\\
    Bandwidth for ED-to-GW link         & $B$ & 125~\si{\kilo\hertz} &~\cite{Casals2017}\\
    Bandwidth for D2D link         & $B_\text{D2D}$ & 250~\si{\kilo\hertz} & \\
    Path loss exponent                  & $\eta$ & 2.7 &~\cite{Georgiou2017}\\
    Carrier frequency                   & $f_c$ & 868~\si{\mega\hertz} &~\cite{Hoeller2018}\\
    Transmission power for LoRa and RT-LoRa
    & \multirow{2}{*}{$P$} & 11~\si{\deci\belmilliwatt} &~\cite{Casals2017}\\
    Transmission power for NCC-LoRa 
    &  & \{0, 11\}~\si{\deci\belmilliwatt} & \\
    Transmission power for D2D link    & $P_\text{D2D}$ & 13~\si{\deci\belmilliwatt} &~\cite{DatasheetLoRa2019}\\
    Noise figure                        & NF & 6~\si{\deci\bel} &~\cite{Hoeller2018}\\
    Sum-interference cancellation threshold & $\delta$ & 6~\si{\deci\bel} &~\cite{Hoeller2018}\\
    D2D receiver sensitivity (FSK)      & $\mathrm{S}_\text{D2D}$ & -82~\si{\deci\belmilliwatt} &~\cite{DatasheetLoRa2019} \\
    D2D link outage probability       & $\mathcal{O}_\mathrm{D2D}$ & 1.2$\%$ & \\
    Transmissions in a time-slot        & $M$ & 2 & \\
    \bottomrule
\end{tabular}%
}\label{tab:sim-parameters}
\end{table}

\subsubsection{NCC-LoRa}
In the ED-to-GW link, the NCC-LoRa scheme presents the same consumption as RT-LoRa in~\eqref{eq:rt-lora-average-current}. However, one needs to also consider the D2D communication in the overall consumption of NCC-LoRa.
Nevertheless, Semtech SX1272 transceiver does not use the same internal circuitry for both the FSK and LoRa modulations~\cite{DatasheetLoRa2019}. Since the D2D phase involves a mutual exchange of information frame data, we consider a D2D model where each cooperating ED performs one transmit and receive operation. The duration of both states depends on selected parameters for the FSK modulation, as defined in the SX1272 transceiver's documentation~\cite{DatasheetLoRa2019}. The values used in our analysis are displayed in Table~\ref{tab:fsk-energy-consumption}, which contains the relevant FSK consumption data extracted from~\cite[Tables~6 and 7]{DatasheetLoRa2019}.

\begin{figure*}[!t]
     \centering
     \begin{subfigure}[b]{.95\columnwidth}
         \centering
         \includegraphics[width=\textwidth]{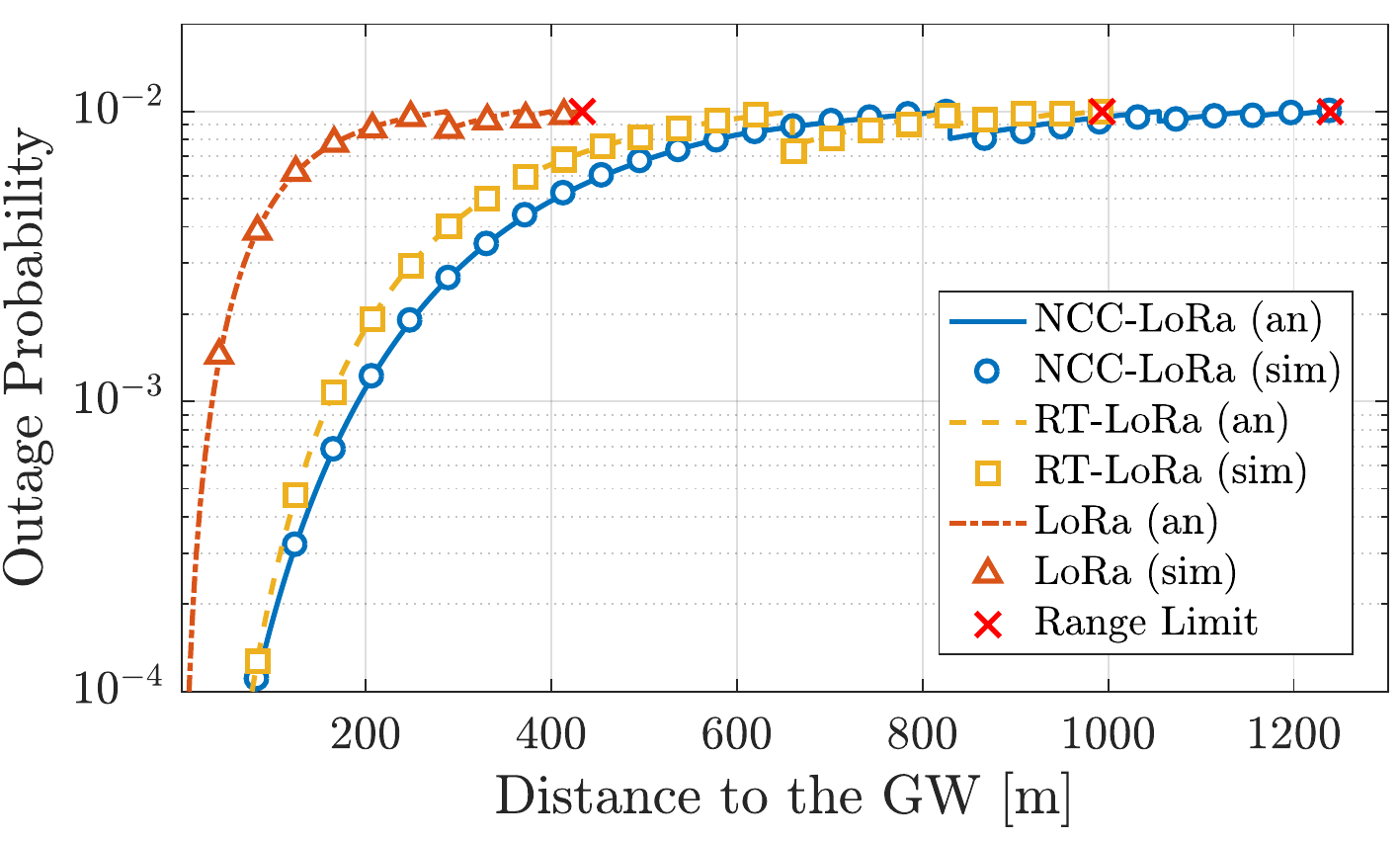}
         \caption{$\rho = 10^{-4}$ EDs/\si{\meter}$^2$}
         \label{fig:Otar1e-2_Out_d1_rho40dB}
     \end{subfigure}
     \hfill
     \begin{subfigure}[b]{.95\columnwidth}
         \centering
         \includegraphics[width=\textwidth]{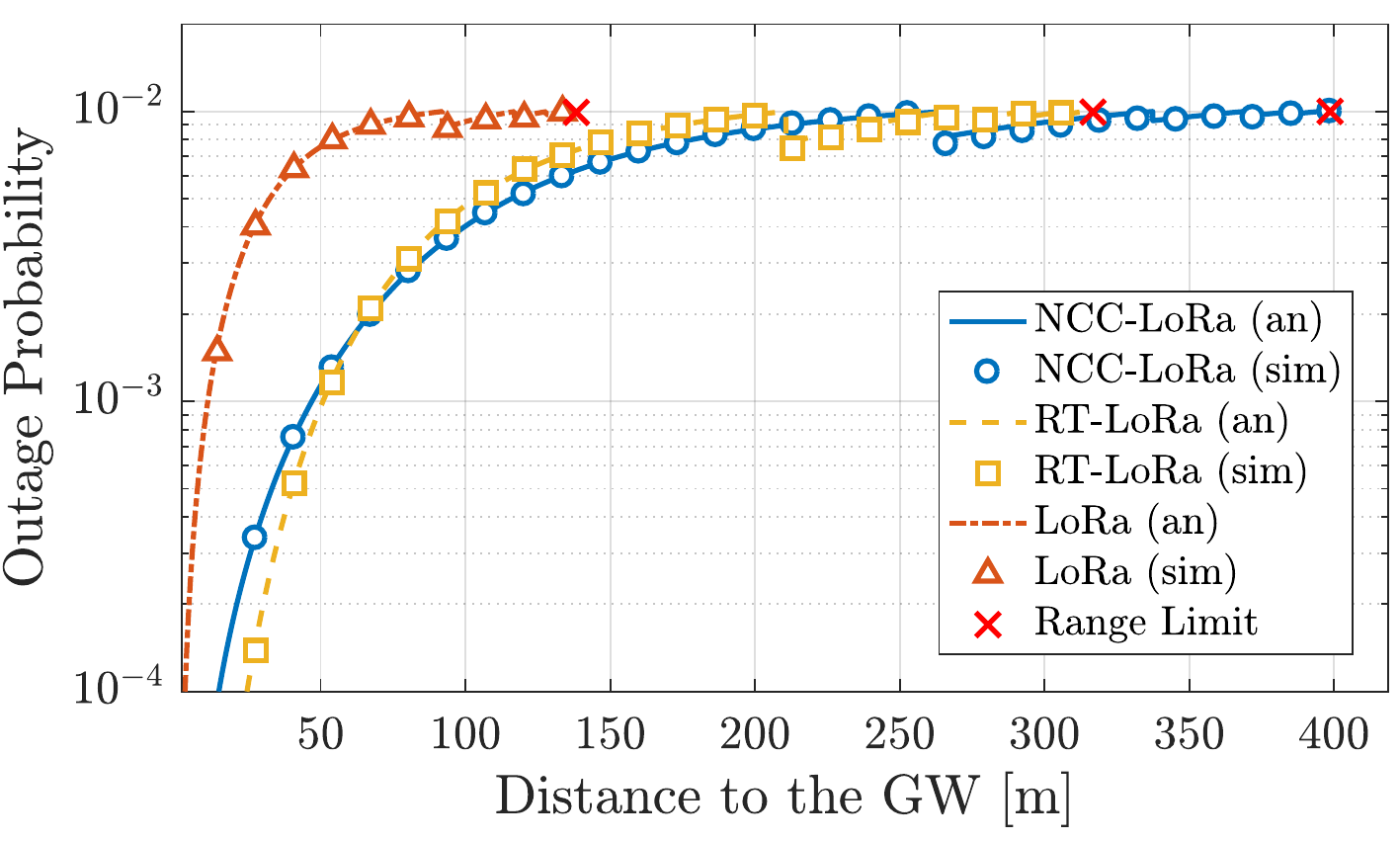}
         \caption{$\rho = 10^{-3}$ EDs/\si{\meter}$^2$}
         \label{fig:Otar1e-2_Out_d1_rho30dB}
     \end{subfigure}
     \hfill
        \caption{Outage probability versus the distance to gateway $d_1$ (\si{\meter}) of the proposed NCC-LoRa (solid line), RT-LoRa (dashed line) and conventional LoRa (dash-dotted line), for $\mathcal{O}_\mathrm{target} = 10^{-2}$ and different density of EDs $\rho$.}
        \label{fig:Out_d1_fixed_Otar1e-2}
\end{figure*}
\begin{figure*}[!t]
     \centering
     \begin{subfigure}[b]{.95\columnwidth}
         \centering
         \includegraphics[width=\textwidth]{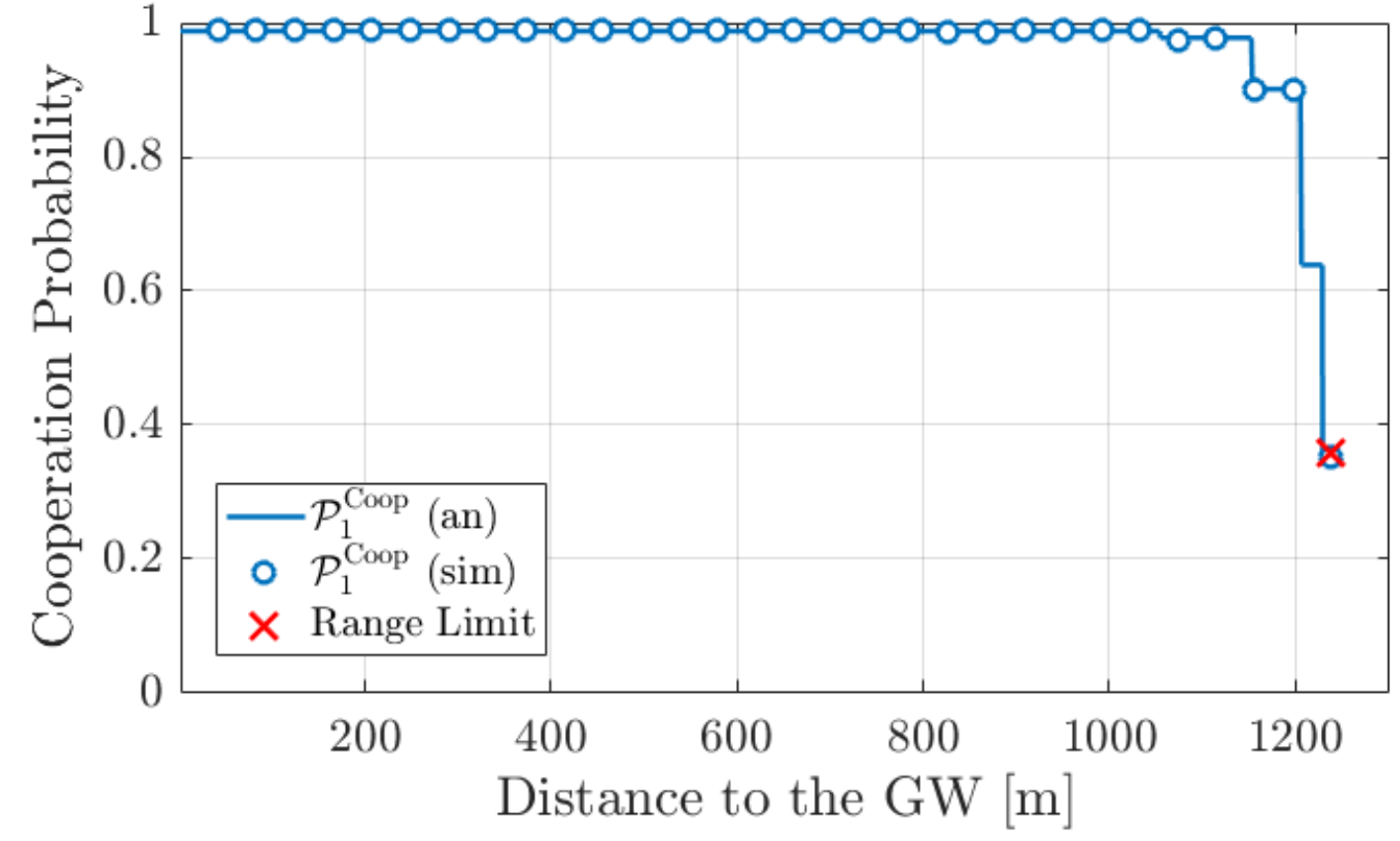}
         \caption{$\rho = 10^{-4}$ EDs/\si{\meter}$^2$}
         \label{fig:Otar1e-2_Pc_d1_rho40dB}
     \end{subfigure}
     \hfill
     \begin{subfigure}[b]{.95\columnwidth}
         \centering
         \includegraphics[width=\textwidth]{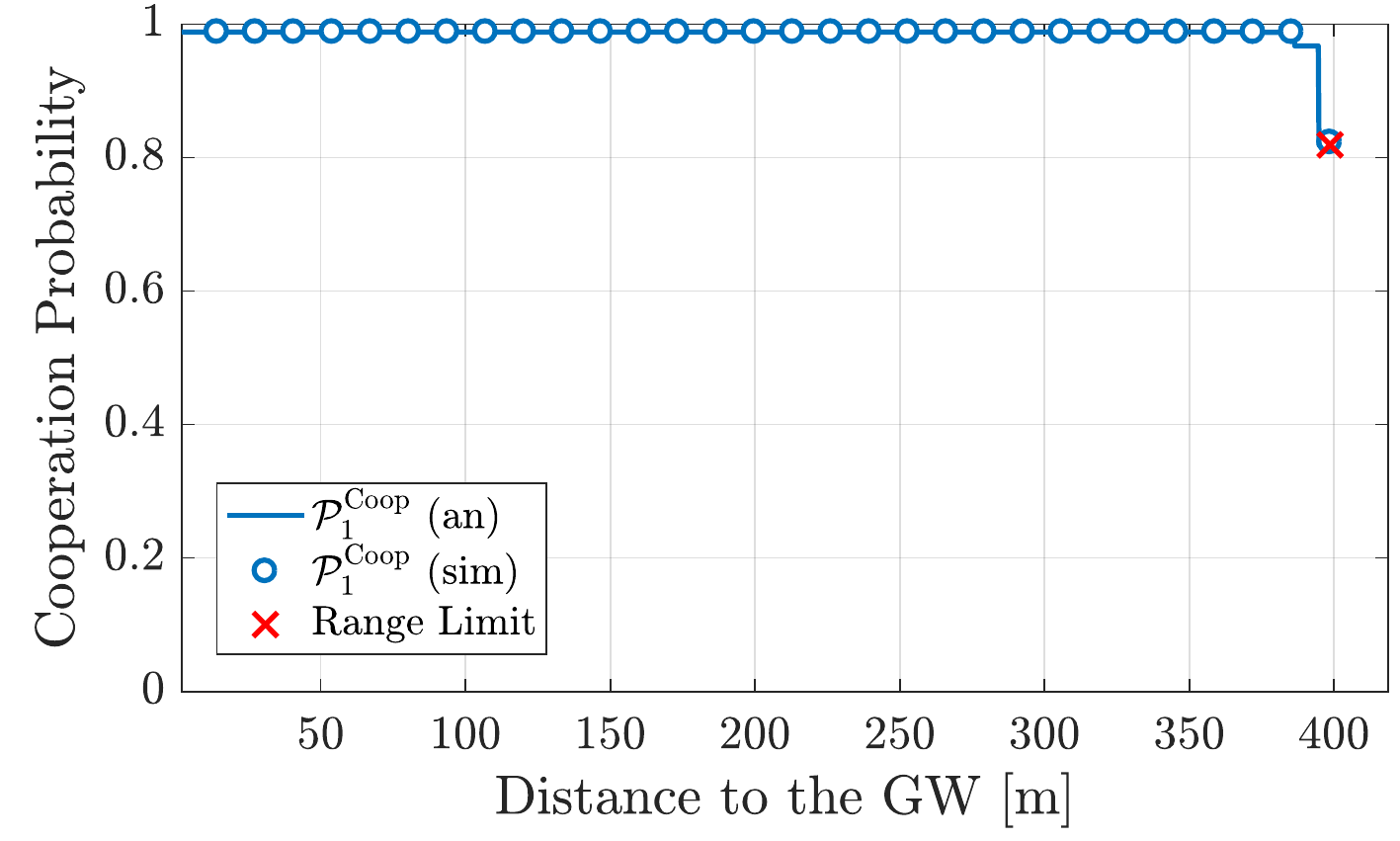}
         \caption{$\rho = 10^{-3}$ EDs/\si{\meter}$^2$}
         \label{fig:Otar1e-2_Pc_d1_rho30dB}
     \end{subfigure}
     \hfill
        \caption{Cooperation probability versus the distance to gateway $d_1$ (\si{\meter}), for $\mathcal{O}_\mathrm{target} = 10^{-2}$ and different density of EDs $\rho$.}
        \label{fig:Pc_d1_fixed_Otar1e-2}
\end{figure*}

Thus, we model the FSK consumption as:
\begin{equation}\label{eq:d2d-average-current}
    I_\mathrm{avg}^\mathrm{D2D} = \frac{1}{\mathrm{T}_\text{slot}}\sum_{i = 1}^{6}\mathrm{T}_i^\text{FSK} I_i^\text{FSK},
\end{equation}
where the FSK values for each state duration $\mathrm{T}_i^\text{FSK}$ and current consumption $I_i^\text{FSK}$ are in Table~\ref{tab:fsk-energy-consumption}.

Finally, the average consumption of NCC-LoRa is
\begin{equation}\label{eq:ncc-lora-average-current}
\begin{split}
    I_\mathrm{avg}^\mathrm{NCC-LoRa} &= \mathcal{P}_\mathrm{neigh}^\mathrm{D2D}\Big[I_\mathrm{avg}^\mathrm{D2D}+I_\mathrm{avg}(2)\Big] 
    \!+\! \Big[1-\mathcal{P}_\mathrm{neigh}^\mathrm{D2D}\Big]I_\mathrm{avg}(2)\\
    & = I_\mathrm{avg}(2) +\underbrace{\mathcal{P}_\mathrm{neigh}^\mathrm{D2D}\, I_\mathrm{avg}^\mathrm{D2D}}_{A}, 
\end{split}
\end{equation}
where term $A$ corresponds to the additional consumption when compared to RT-LoRa, when both schemes operate with the same SF. However, since NCC-LoRa has an improved reliability than RT-LoRa, one could expect the former to achieve the same communication range as the latter with a smaller SF, and thus consuming less energy in the ED-to-GW link. We evaluate in the next section whether this non-trivial relationship between the savings in the ED-to-GW link and the additional consumption imposed by the D2D communication leads to a positive final balance.
\section{Numerical Results} \label{sec:numerical-results}

In this section we present some numerical results aiming at comparing the performance of the proposed NCC-LoRa scheme to RT-LoRa and conventional LoRa, as well as to support the analysis developed throughout the paper. We resort to the Monte Carlo method to obtain the numerical results, by averaging the outcome of $10^6$ independent trials. Unless stated otherwise, we adopt the parameters in Table~\ref{tab:sim-parameters}\footnote{The value $P\!=\!11$~\si{\deci\belmilliwatt} follows~\cite{SantAna2020,Casals2017}. A different transmission power could be used without altering the main conclusions of this work, provided that other parameters comply with Table~\ref{tab:sim-parameters}.}. 
In the figures, ``(an)''  and ``(sim)'' refers respectively to the analytical and numerical results.

Note that FSK mode of Semtech's SX1272 transceiver operates with a sensitivity $\mathrm{S}_\text{D2D} = -92$~\si{\deci\belmilliwatt} for $B = 250$~\si{\kilo\hertz} and Bit Error Rate (BER) of 0.1$\%$~\cite[Table 8]{DatasheetLoRa2019}. However, assuming that an outage occurs upon receiving at least a single erroneous bit, a $N_b\! =\! 120$ bit message ($\text{PL}\! =\! 9$ bytes from the payload plus $6$ additional bytes for the FSK fixed payload length frame formatting~\cite{DatasheetLoRa2019}) would experience an outage probability of $\mathcal{O}_\mathrm{D2D} = 11.3\%$. Following the performance of FSK under Rayleigh fading~\cite{Goldsmith2005}, we add 10~\si{dB} to the receiver sensitivity from~\cite[Table 8]{DatasheetLoRa2019} ($\mathrm{S}_\text{D2D} = -92 + 10 = -82$~\si{\deci\belmilliwatt}), which results in $\text{BER} \approx 10^{-4}$, leading to $\mathcal{O}_\mathrm{D2D} = 1.2\%$, as shown in Table~\ref{tab:sim-parameters}. 

\subsection{Outage Performance} \label{ssec:outage-performance}
Fig.~\ref{fig:Out_d1_fixed_Otar1e-2} and Fig.~\ref{fig:Pc_d1_fixed_Otar1e-2} present respectively the outage probability and the cooperation probability as a function of the communication range, both for $\rho = \{10^{-4},10^{-3}\}$~EDs/\si{\meter}$^2$. We adopt the SF allocation procedure discussed in Section~\ref{sssec:sf_allocation}, where the boundaries of each SF are determined by a given maximum allowed outage probability, which has been set as $\mathcal{O}_\mathrm{target} = 10^{-2}$ in Fig.~\ref{fig:Out_d1_fixed_Otar1e-2} and Fig.~\ref{fig:Pc_d1_fixed_Otar1e-2}, and that depends on both capture and connection probabilities. In Fig.~\ref{fig:Otar1e-2_Out_d1_rho40dB}, even with the relatively low density of $\rho = 10^{-4}$ EDs/\si{\meter}$^2$, NCC-LoRa already delivers an improvement over RT-LoRa: the gain in the number of supported EDs in the network, calculated as $N_\mathrm{sup} = \rho \, \pi \mathcal{R}^2$, is of approximately 55.5$\%$. This indicates that the NCC mode was capable of reducing the impact of collisions in the ED-to-GW link. We can also extrapolate from the maximum range results (roughly 993~\si{\meter} for RT-LoRa and 1239~\si{\meter} for NCC-LoRa) that the EDs under NCC-LoRa were rarely having to fall back to the RT-LoRa option. Indeed, cooperation could be established with high frequency, as depicted in Fig.~\ref{fig:Otar1e-2_Pc_d1_rho40dB}. However, Fig.~\ref{fig:Otar1e-2_Pc_d1_rho40dB} also shows that, due to a narrower SF coverage region of $\xi\left(\mathcal{S}_\mathrm{F}\right)<100~\si{\meter}$ for $\mathcal{S}_\mathrm{F}\geq 9$, the probability of cooperation declines at the edges of the network. By increasing the network density by a factor of 10, and as a consequence of the resulting higher cooperation probability at the network edge, as displayed in Fig.~\ref{fig:Otar1e-2_Pc_d1_rho30dB}, the NCC-LoRA performance, presented in Fig.~\ref{fig:Otar1e-2_Out_d1_rho30dB}, elevated the $N_\mathrm{sup}$ gain to 58.5$\%$ when compared to RT-LoRa.

\begin{figure}[!t]
\center
 \includegraphics[width=.95\columnwidth]{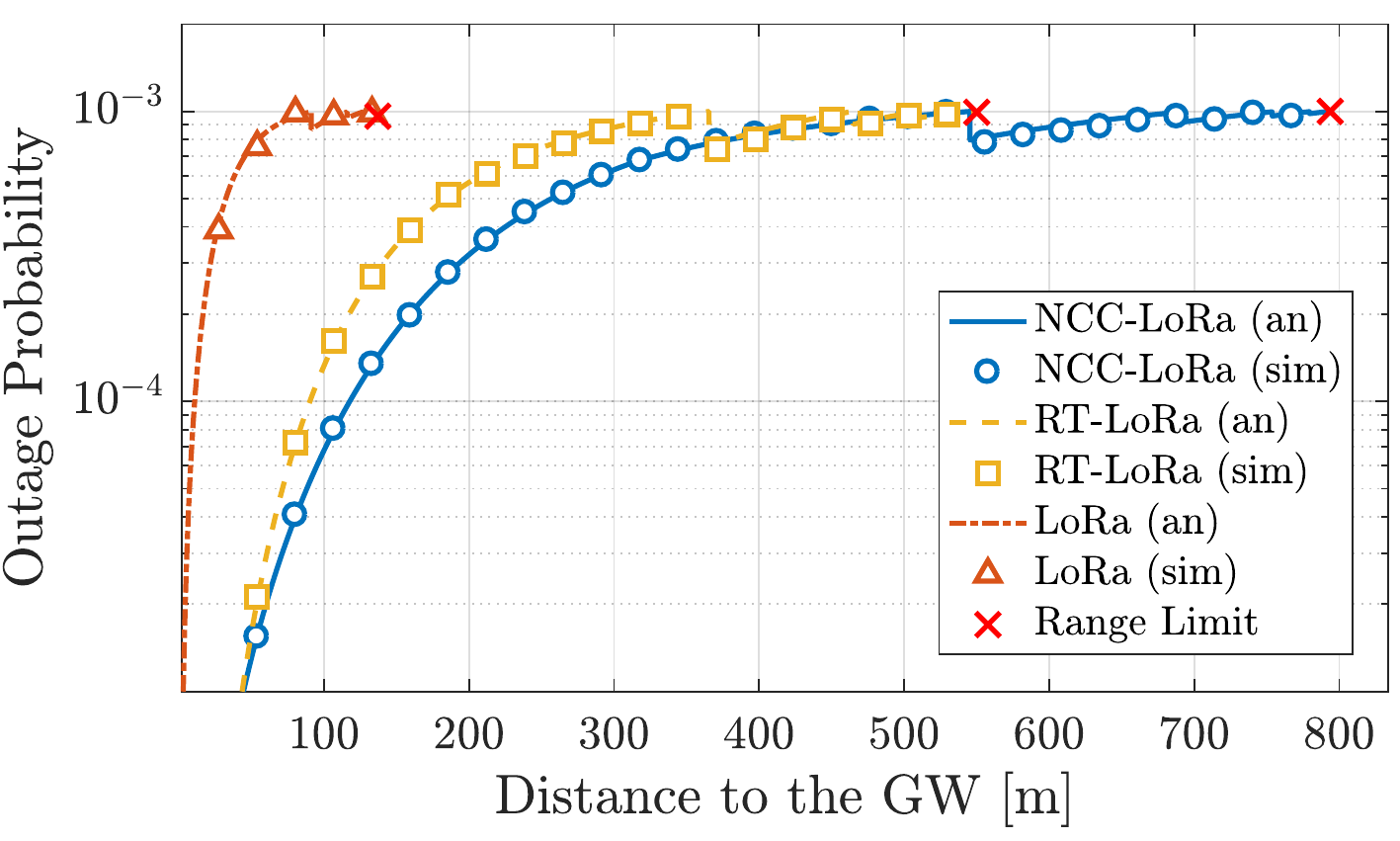}\label{fig:Otar1e-3_Out_d1_rho30dB}
\caption{Outage probability versus the distance to gateway $d_1$ (\si{\meter}), for $\mathcal{O}_\mathrm{target} = 10^{-3}$ and $\rho = 10^{-4}$ EDs/\si{\meter}$^2$.}
\label{fig:Out_d1_fixed_Otar1e-3}
\end{figure}
\begin{figure}[!t]
    \centering
    \includegraphics[width=.95\columnwidth]{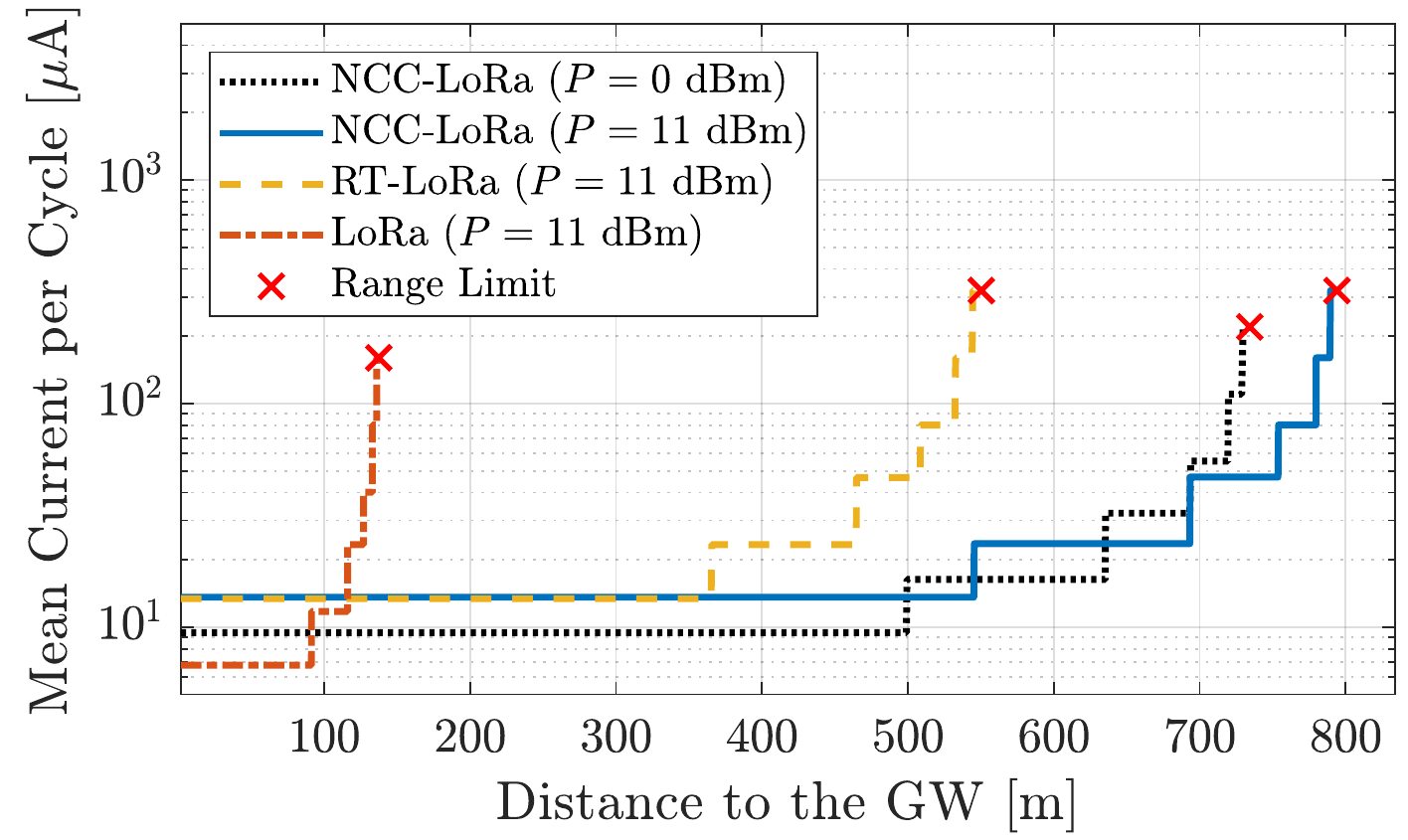}
    \caption{Average current consumption versus the distance to gateway $d_1$ (\si{\meter}) for $\rho = 10^{-4}$ EDs/\si{\meter}$^2$ and $\mathcal{O}_\mathrm{target}=10^{-3}$.}
    \label{fig:Consumption_d1}
\end{figure}

Furthermore, it can be seen in Fig.~\ref{fig:Pc_d1_fixed_Otar1e-2} that the approximation from~\eqref{eq:Ac-approx} is useful in obtaining a good estimate on the value of the cooperation area, particularly in the proximity of the SF boundaries. However, we find it relevant to mention that as the difference between the SF coverage region and cooperation distance increases, \textit{i.e.} for $\xi(\mathcal{S}_\mathrm{F}) \gg d_1^\mathrm{Coop}$, the proposed approximation in~\eqref{eq:Ac-approx} becomes more imprecise in modelling the outage probability for EDs not at the SF boundaries. Nevertheless, one can see from Fig.~\ref{fig:Out_d1_fixed_Otar1e-2} that any inaccuracy in~\eqref{eq:Ac-approx} has a small impact in the overall analytical outage probability of the NCC-LoRa scheme. 

In a scenario subject to a more strict outage requirement of $\mathcal{O}_\mathrm{target}~=~10^{-3}$ in Fig.~\ref{fig:Out_d1_fixed_Otar1e-3}, even though the number of supported EDs is reduced to all schemes, the gain of NCC-LoRa over RT-LoRa is increased to 120.5$\%$. It is also noteworthy that both RT-LoRa and NCC-LoRa considerably outperform the conventional LoRa implementation on all studied scenarios. 
\subsection{Energy Performance}\label{ssec:energy-performance}

Fig.~\ref{fig:Consumption_d1} compares the average current consumption of  NCC-LoRa to that of RT-LoRa and conventional LoRa. For a fixed SF and using the same transmission power $P = 11~\si{\deci\belmilliwatt}$ in the ED-to-GW link, NCC-LoRa presents the highest consumption (due to the additional D2D phase), while LoRa has the lowest consumption. However, when adopting NCC-LoRa, one can extend the upper boundary of each SF while maintaining the reliability requirement due to the improved outage probability performance provided by NCC. Thus, when the distance to the GW increases, instead of adopting higher SFs (and consequently increasing the energy consumption) with the same high pace as in conventional LoRa, NCC-LoRa can achieve the same performance generally with a lower SF, consuming less energy even when taking the additional consumption of the D2D communication into account. This can be seen in Fig.~\ref{fig:Consumption_d1} for $d_1 > 365$~\si{\meter}, where NCC-LoRa becomes the most energy efficient scheme. 

One can also see from Fig.~\ref{fig:Consumption_d1} that NCC-LoRa outperforms the communication range of RT-LoRa even when the former operates with a reduced transmission power $P~=~0~\si{\deci\belmilliwatt}$ in the Ed-to-GW link. In this scenario, due to the reduced transmission power, the consumption of NCC-LoRa is also lower than that of RT-LoRa. Thus, the additional consumption of the D2D phase could be compensated by reducing $P$. As a drawback of reducing $P$, one has that the communication range $\mathcal{R}$ is reduced $7.5\%$ in this scenario.

Fig.~\ref{fig:Consumption_d1} also depicts an interesting tradeoff inherent to NCC-LoRa: for some values of distances to the GW (for example, $500 \leq d_1 \leq 545$ m), the consumption of NCC-LoRa is lower when adopting $P=11$ dBm than when setting $P=0$ dBm. This is due to the fact that the increased transmission power makes it possible to adopt a smaller SF, which reduces the ToA and consequently leads to a reduced consumption.

\section{Final Comments} \label{sec:final-comments}

In this paper we propose NCC-LoRa, where the EDs are capable of exchanging messages using D2D communication, as well as to transmitting linear combinations of more than one frame to a common GW. In our analysis, we consider both connection and collision probabilities when modelling the outage probability of an ED, as well as a realistic power consumption model. Our results indicate that the proposed NCC-LoRa scheme provides considerable gains in terms of outage probability while possibly consuming less energy than both conventional LoRa and RT-LoRa, even when considering the additional D2D communication. Our analysis was based on technical specifications of Semtech's SX1272 transceiver~\cite{DatasheetLoRa2019}. Finally, the empirical validation of the consumption of NCC-LoRa is left as a future work.

\appendices
\section{Proof of Lemma~\ref{lem:cooperation-area-approx}}\label{ap:cooperation-area-approx}
Given that cooperation can only be established between EDs using the same SF, the cooperation area depends on the relationship between the width $\xi(\mathcal{S}_\mathrm{F})$ and the cooperation distance $d_1^\mathrm{Coop}$ in Definition~\ref{def:cooperation-distance}, yielding two different scenarios:

\begin{figure}[!t]
     \centering
     \begin{subfigure}[b]{0.48\columnwidth}
         \centering
         \includegraphics[width=\textwidth]{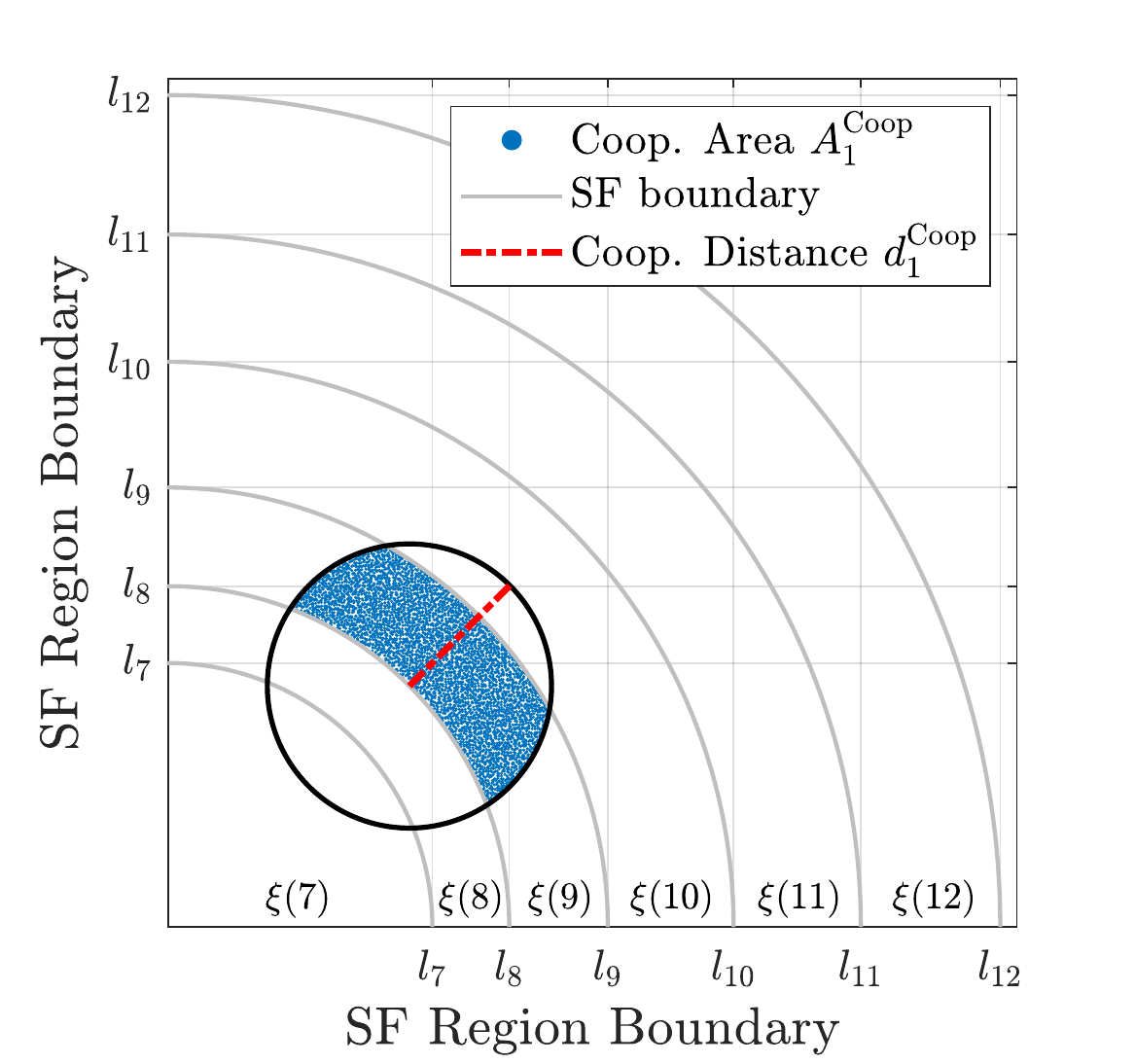}
         \caption{$\xi(\mathcal{S}_\mathrm{F}) < d_1^\mathrm{Coop}$}
         \label{fig:Ac-approx-a}
     \end{subfigure}
     \hfill
     \begin{subfigure}[b]{0.48\columnwidth}
         \centering
         \includegraphics[width=\textwidth]{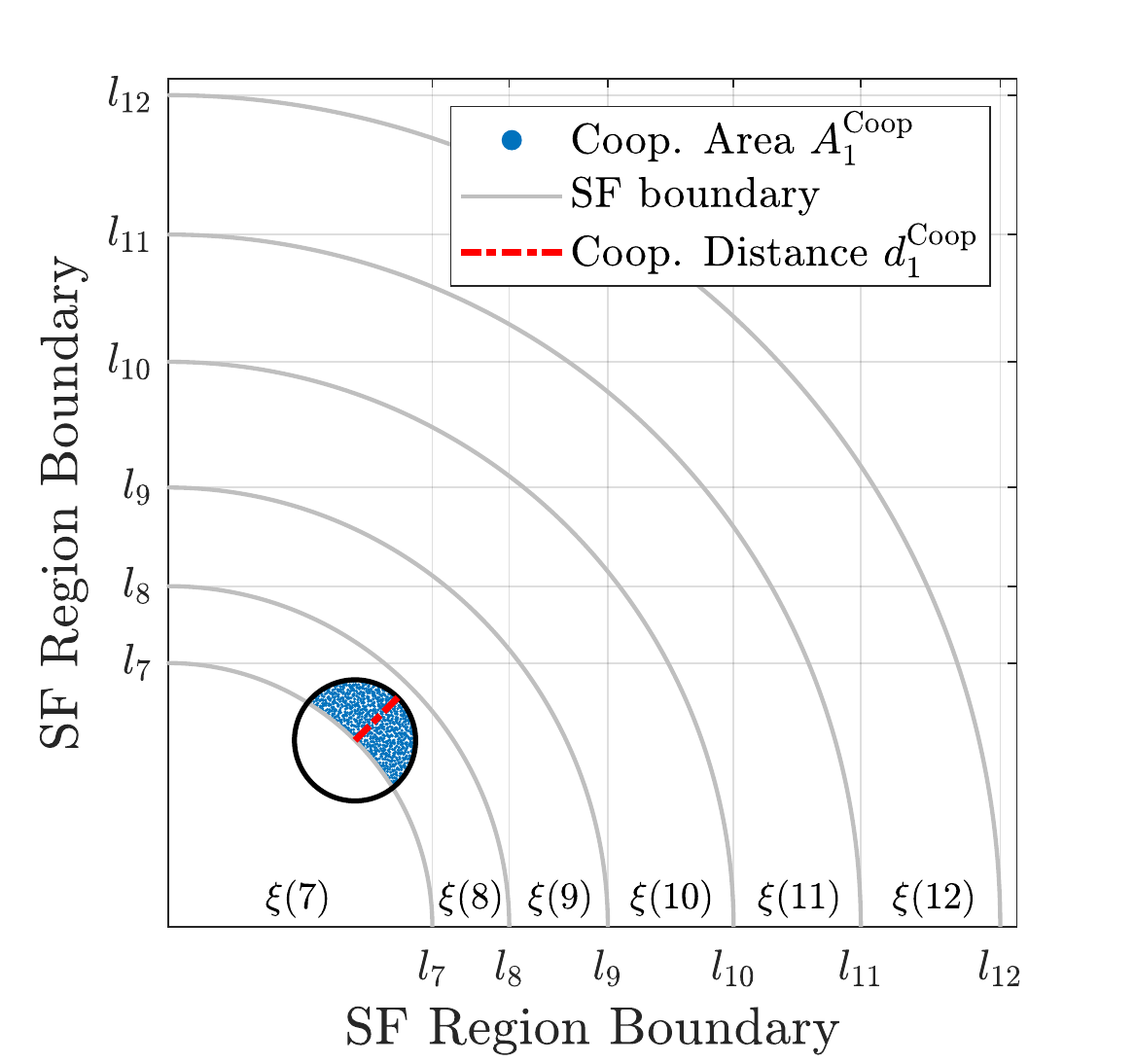}
         \caption{$\xi(\mathcal{S}_\mathrm{F}) \geq d_1^\mathrm{Coop}$}
         \label{fig:Ac-approx-b}
     \end{subfigure}
     \hfill
        \caption{Possible scenarios for the cooperation area.}
        \label{fig:Ac-approx}
\end{figure}
\subsubsection{$\xi(\mathcal{S}_\mathrm{F}) < d_1^\mathrm{Coop}$} This scenario is illustrated in Fig.~\ref{fig:Ac-approx-a}, and corresponds to the situation where $d_1^\mathrm{Coop}$ can be approximated by a rectangle of width $2 d_1^\mathrm{Coop}$ and height $\xi(\mathcal{S}_\mathrm{F})$, {\it i.e.} 
\begin{equation}\label{eq:Ac-approx-cond1}
    A_{1a,\mathrm{approx}}^\mathrm{Coop} \approx 2 d_1^\mathrm{Coop}\xi(\mathcal{S}_\mathrm{F}). 
\end{equation}

\subsubsection{$\xi(\mathcal{S}_\mathrm{F}) \geq d_1^\mathrm{Coop}$} Under this circumstance, depicted in Fig.~\ref{fig:Ac-approx-b}, we consider the worst case where the ED is located at the edge of the SF, and approximate the cooperation area as the area of a semicircle with radius $d_1^\mathrm{Coop}$, {\it i.e.}  
\begin{equation}\label{eq:Ac-approx-cond2}
    A_{1b,\mathrm{approx}}^\mathrm{Coop} \approx \frac{\pi}{2} \left(d_1^\mathrm{Coop}\right)^2. 
\end{equation}

Finally, the approximated value~\eqref{eq:Ac-approx} is obtained in a conservative way as the minimum between \eqref{eq:Ac-approx-cond1} and~\eqref{eq:Ac-approx-cond2}. 



\begin{thebibliography}{10}
\providecommand{\url}[1]{#1}
\csname url@samestyle\endcsname
\providecommand{\newblock}{\relax}
\providecommand{\bibinfo}[2]{#2}
\providecommand{\BIBentrySTDinterwordspacing}{\spaceskip=0pt\relax}
\providecommand{\BIBentryALTinterwordstretchfactor}{4}
\providecommand{\BIBentryALTinterwordspacing}{\spaceskip=\fontdimen2\font plus
\BIBentryALTinterwordstretchfactor\fontdimen3\font minus
  \fontdimen4\font\relax}
\providecommand{\BIBforeignlanguage}[2]{{%
\expandafter\ifx\csname l@#1\endcsname\relax
\typeout{** WARNING: IEEEtran.bst: No hyphenation pattern has been}%
\typeout{** loaded for the language `#1'. Using the pattern for}%
\typeout{** the default language instead.}%
\else
\language=\csname l@#1\endcsname
\fi
#2}}
\providecommand{\BIBdecl}{\relax}
\BIBdecl

\bibitem{Centenaro2016}
M.~{Centenaro}, L.~{Vangelista}, A.~{Zanella}, and M.~{Zorzi}, ``{Long-Range
  Communications in Unlicensed Bands: the Rising Stars in the IoT and Smart
  City Scenarios},'' \emph{{IEEE} Wireless Commun.}, vol.~23, no.~5, pp.
  60--67, Oct. 2016.

\bibitem{Raza2017}
U.~{Raza}, P.~{Kulkarni}, and M.~{Sooriyabandara}, ``{Low Power Wide Area
  Networks: An Overview},'' \emph{{IEEE} Commun. Surveys Tuts.}, vol.~19,
  no.~2, pp. 855--873, Secondquarter 2017.

\bibitem{Ikpehai2019}
A.~{Ikpehai}, B.~{Adebisi}, K.~M. {Rabie}, K.~{Anoh}, R.~E. {Ande},
  M.~{Hammoudeh}, H.~{Gacanin}, and U.~M. {Mbanaso}, ``{Low-Power Wide Area
  Network Technologies for Internet-of-Things: A Comparative Review},''
  \emph{{IEEE} Internet Things J.}, vol.~6, no.~2, pp. 2225--2240, Apr. 2019.

\bibitem{Sundaram2020}
J.~P. {Shanmuga Sundaram}, W.~{Du}, and Z.~{Zhao}, ``{A Survey on {L}o{R}a
  Networking: Research Problems, Current Solutions, and Open Issues},''
  \emph{{IEEE} Commun. Surveys Tuts.}, vol.~22, no.~1, pp. 371--388,
  Firstquarter 2020.

\bibitem{LoRaPATENT}
O.~B.~A. Seller, ``{Wireless Communication Method},'' U.S. Patent 9 647 718,
  May 09, 2017.

\bibitem{LoRaWAN}
\emph{LoRaWAN\textsuperscript{TM} 1.1 Specification}, LoRa Alliance, Inc.

\bibitem{Georgiou2017}
O.~{Georgiou} and U.~{Raza}, ``{Low {P}ower {W}ide {A}rea {N}etwork Analysis:
  Can {L}o{R}a Scale?}'' \emph{{IEEE} Wireless Commun. Lett.}, vol.~6, no.~2,
  pp. 162--165, Apr. 2017.

\bibitem{Mahmood2019}
A.~{Mahmood}, E.~{Sisinni}, L.~{Guntupalli}, R.~{Rond{\'o}n}, S.~A. {Hassan},
  and M.~{Gidlund}, ``{Scalability Analysis of a {L}o{R}a Network Under
  Imperfect Orthogonality},'' \emph{{IEEE} Trans. Ind. Informat.}, vol.~15,
  no.~3, pp. 1425--1436, Mar. 2019.

\bibitem{Hoeller2018}
A.~{Hoeller}, R.~D. {Souza}, O.~L. {Alcaraz L\'{o}pez}, H.~{Alves}, M.~{de
  Noronha Neto}, and G.~{Brante}, ``{Analysis and Performance Optimization of
  {L}o{R}a Networks With Time and Antenna Diversity},'' \emph{IEEE Access},
  vol.~6, pp. 32\,820--32\,829, Jun. 2018.

\bibitem{SantAna2020}
J.~Sant'Ana, A.~Hoeller~Jr, R.~Souza, S.~Montejo~S\'anchez, H.~Alves, and
  M.~Neto, ``{Hybrid Coded Replication in LoRa Networks},'' \emph{{IEEE} Trans.
  Ind. Informat.}, vol.~16, no.~8, pp. 5577--5585, Aug. 2020.

\bibitem{SantAna2020SIC}
J.~M. {de Souza Sant’Ana}, A.~{Hoeller}, R.~D. {Souza}, H.~{Alves}, and
  S.~{Montejo-Sánchez}, ``{LoRa Performance Analysis with Superposed Signal
  Decoding},'' \emph{{IEEE} Wireless Commun. Lett.}, vol.~9, no.~11, pp.
  1865--1868, Nov. 2020.

\bibitem{Jiang2021}
X.~{Jiang}, H.~{Zhang}, E.~A. {Barsallo Yi}, N.~{Raghunathan}, C.~{Mousoulis},
  S.~{Chaterji}, D.~{Peroulis}, A.~{Shakouri}, and S.~{Bagchi}, ``{Hybrid
  Low-Power Wide-Area Mesh Network for IoT Applications},'' \emph{{IEEE}
  Internet Things J.}, vol.~8, no.~2, pp. 901--915, Jan. 2021.

\bibitem{Hsu2018}
S.~{Hsu}, C.~{Lin}, C.~{Wang}, and W.~{Chen}, ``{Breaking Bandwidth Limitation
  for Mission-Critical {I}o{T} Using Semisequential Multiple Relays},''
  \emph{{IEEE} Internet Things J.}, vol.~5, no.~5, pp. 3316--3329, Oct. 2018.

\bibitem{Borkotoky2019}
S.~S. {Borkotoky}, U.~{Schilcher}, and C.~{Bettstetter}, ``{Cooperative
  Relaying in {L}o{R}a Sensor Networks},'' in \emph{IEEE Global Communications
  Conference (GLOBECOM)}, Dec. 2019, pp. 1--5.

\bibitem{Mikhaylov2017}
K.~{Mikhaylov}, J.~{Pet{\"a}j{\"a}j{\"a}rvi}, J.~{Haapola}, and A.~{Pouttu},
  ``{{D2D} communications in {L}o{R}a{WAN} {L}ow {P}ower {W}ide {A}rea
  {N}etwork: From idea to empirical validation},'' in \emph{IEEE International
  Conference on Communications Workshops (ICC Workshops)}, May 2017, pp.
  737--742.

\bibitem{Kim2018}
J.~{Kim} and J.~{Song}, ``{A Secure Device-to-Device Link Establishment Scheme
  for {L}o{R}a{WAN}},'' \emph{{IEEE} Sensors J.}, vol.~18, no.~5, pp.
  2153--2160, Mar. 2018.

\bibitem{Laneman2004}
J.~N. Laneman, D.~N.~C. Tse, and G.~W. Wornell, ``{Cooperative Diversity in
  Wireless Networks: Efficient Protocols and Outage Bahavior},'' \emph{{IEEE}
  Trans. Inf. Theory}, vol.~50, no.~12, pp. 3062--3080, Dec. 2004.

\bibitem{Ahlswede2000}
R.~Ahlswede, N.~Cai, S.-Y. Li, and R.~Yeung, ``{Network Information Flow},''
  \emph{{IEEE} Trans. Inf. Theory}, vol.~46, no.~4, pp. 1204 -- 1216, 2000.

\bibitem{Xiao2010}
M.~Xiao and M.~Skoglund, ``{Multiple-User Cooperative Communications Based on
  Linear Network Coding},'' \emph{{IEEE} Trans. Commun.}, vol.~58, no.~12, pp.
  3345--3351, Dec. 2010.

\bibitem{Rebelatto2012}
J.~L. Rebelatto, B.~F. Uch{\^o}a-Filho, Y.~Li, and B.~Vucetic, ``{Multi-User
  Cooperative Diversity through Network Coding Based on Classical Coding
  Theory},'' \emph{{IEEE} Trans. Signal Process.}, vol.~60, no.~2, pp.
  916--926, Feb. 2012.

\bibitem{Wu2015}
Y.~Wu, W.~Liu, S.~Wang, W.~Guo, and X.~Chu, ``{Network Coding in
  Device-to-Device (D2D) Communications Underlaying Cellular Networks},'' in
  \emph{IEEE International Conference on Communications (ICC)}, 2015, pp.
  2072--2077.

\bibitem{Montejo2019}
S.~{Montejo-S\'anchez}, C.~A. {Azurdia-Meza}, R.~D. {Souza}, E.~M.~G.
  {Fernandez}, I.~{Soto}, and A.~{Hoeller}, ``{Coded Redundant Message
  Transmission Schemes for Low-Power Wide Area {I}o{T} Applications},''
  \emph{{IEEE} Wireless Commun. Lett.}, vol.~8, no.~2, pp. 584--587, Apr. 2019.

\bibitem{ANT}
\BIBentryALTinterwordspacing
\emph{{ANT Message Protocol and Usage}}, Dynastream Innovations Inc., 2014,
  rev. 5.1. [Online]. Available:
  \url{thisisant.com/resources/ant-message-protocol-and-usage}
\BIBentrySTDinterwordspacing

\bibitem{DatasheetLoRa2019}
\emph{{{SX1272/73} - 860 {M}Hz to 1020 {M}Hz Low Power Long Range
  Transceiver}}, Semtech Corporation, Wireless \& Sensing Products, Jan. 2019,
  rev. 4.

\bibitem{Goldsmith2005}
A.~Goldsmith, \emph{{Wireless Communications}}.\hskip 1em plus 0.5em minus
  0.4em\relax Cambridge University Press, 2005.

\bibitem{Cuomo2017}
F.~Cuomo, M.~Campo, A.~Caponi, G.~Bianchi, G.~Rossini, and P.~Pisani,
  ``{EXPLoRa: Extending the Performance of {LoRa} by Suitable Spreading Factor
  Allocations},'' in \emph{{IEEE} {International} {Conference} on {Wireless}
  and {Mobile} {Computing}, {Networking} and {Communications} ({WiMob})}, Oct.
  2017, pp. 1--8.

\bibitem{Berioli2016}
M.~Berioli, G.~Cocco, G.~Liva, and A.~Munari, ``{Modern Random Access
  Protocols},'' \emph{Found. and Trends in Netw.}, vol.~10, no.~4, pp.
  317--446, 2016.

\bibitem{Daalhuis2010}
A.~B.~O. Daalhuis, ``{Hypergeometric Function},'' in \emph{NIST Handbook of
  Mathematical Functions}, 1st~ed., F.~W.~J. Olver, D.~W. Lozier, R.~F.
  Boisvert, and C.~W. Clark, Eds.\hskip 1em plus 0.5em minus 0.4em\relax New
  York, NY, USA: Cambridge Univ. Press, 2010, ch.~15, pp. 383--402.

\bibitem{BLE_NDP}
B.~{Luo}, Y.~{Yao}, and Z.~{Sun}, ``{Performance Analysis Models of BLE
  Neighbor Discovery: A Survey},'' \emph{{IEEE} Internet Things J.}, vol.~8,
  no.~11, pp. 8734--8746, Jun. 2021.

\bibitem{Daley2013}
D.~Daley and D.~Vere-Jones, \emph{{An Introduction to the Theory of Point
  Processes}}, ser. Springer Series in Statistics.\hskip 1em plus 0.5em minus
  0.4em\relax Springer New York, 2013.

\bibitem{Broch1998}
J.~{Broch}, D.~A. {Maltz}, D.~B. {Johnson}, Y.-C. {Hu}, and J.~{Jetcheva}, ``A
  performance comparison of multi-hop wireless ad hoc network routing
  protocols,'' in \emph{Proceedings of the 4th annual ACM/IEEE international
  conference on Mobile computing and networking}, 1998, pp. 85--97.

\bibitem{LoRaCalculator}
\BIBentryALTinterwordspacing
{Semtech Corp.}, ``{SX}1272 {L}o{R}a {C}alculator.'' [Online]. Available:
  \url{https://semtech.my.salesforce.com/sfc/p/#E0000000JelG/a/2R000000HUhK/6T9Vdb3_ldnElA8drIbPYjs1wBbhlWUXej8ZMXtZXOM}
\BIBentrySTDinterwordspacing

\bibitem{Casals2017}
L.~Casals, B.~Mir, R.~Vidal, and C.~Gomez, ``{Modeling the Energy Performance
  of {L}o{R}a{WAN}},'' \emph{Sensors}, vol.~17, no.~10, Oct. 2017.

\end{thebibliography}
\end{document}